\pdfoutput=1
\documentstyle[12pt,graphicx,natbib]{article}
\def\sun{\hbox{$\odot$}}
\def\aap{A\&A\,  }
\def\aaps{A\&AS  }
\def\aj{AJ  }
\def\apj{ApJ\,  }
\def\apjl{ApJ\,  }
\def\apjs{ApJS  }
\def\apss{Astrophysics and Space Science  }
\def\cjaa{Chinese J. Astron. Astrophys.  }
 
\def\mnras{MNRAS\,  }
\def\nat{Nature\,  }

\def\physrep{Phys. Rep.\,  }
\def\pre{Phys. Rev. E   }
\def\physa{Phys. A    }
\def\rmp{Rev. Mod. Phys.  }
\def\za{Z. Astrophys.  } 

\begin{document}
\title     
{
A geometrical model for the
catalogs of galaxies
}
\author
{L. Zaninetti             \\
Dipartimento di Fisica Generale, \\
           Via Pietro Giuria 1,   \\
           10125 Torino, Italy
}

\maketitle
\section*{}
The  3D network 
originated by  the faces of  irregular 
Poissonian Voronoi Polyhedrons
may represent the backbone on which the galaxies are 
originated.
As a consequence  the spatial appearance 
of the catalogs of galaxies can be reproduced.
The selected catalogs to simulate are
the 2dF Galaxy Redshift Survey
and the  Third Reference Catalog of Bright Galaxies.
In order to explain the number of observed galaxies
for a given flux/magnitude as a function of the
redshift, 
 the photometric properties 
of the galaxies should be carefully examined  
from both the astronomical and 
theoretical point of view. 
The statistics of the 
Voronoi  normalized volume is modeled
by two distributions and 
 the Eridanus super-void is identified as the largest
volume belonging to the Voronoi Polyhedron.
The behavior of the correlation function for galaxies is 
simulated by 
adopting the framework of  thick faces of  Voronoi
Polyhedrons on  short scales,
while  adopting  standard arguments on  large scales.

\section{Introduction}

During the last thirty years the spatial distribution of 
galaxies has been   investigated  from 
the point of view of 
geometrical  and  physical theories.
One first target was to reproduce the two-point 
correlation function $\xi (r)$ for galaxies which 
on  average scales  as $\approx (\frac {r} {5.7Mpc})^{-1.8}$,
see 
\citet{Jones2005} 
and 
\citet{Gallagher2000}.
The statistical theories  
of spatial galaxy distribution
can be classified as 
\begin{itemize}
\item {\bf Levy flights}: the random walk with a variable 
 step  length    can lead to a correlation function in  agreement
with the observed data, see 
\citet{Mandelbrot1975b},
\citet{Soneira1977},
\citet{Soneira1978} and
\citet{Peebles1980}.
\item {\bf Percolation}: the theory of 
          primordial explosions
         can lead to the formation of structures, see 
\citet{Charlton1986} 
and
\citet{ferraro}. 
          Percolation is also used 
         as a tool to organize : (i) the mass and 
         galaxy distributions obtained in 3D
         simulations of  cold dark matter (CDM)
         and  hot dark matter (HDM), see 
\citet{Klypin1993}, 
         (ii)
         the galaxy groups  and clusters in 
         volume-limited samples of the
         Sloan Digital Sky Survey (SDSS), see 
         \citet{Berlind2006}
\end{itemize}
The geometrical models are well represented
by the {\bf Voronoi Diagrams}. 
         The applications to galaxies 
         started with
         \citet{icke1987}, 
         where a sequential clustering
         process was adopted in order to insert the initial seeds,
         and they continued 
         with 
         \citet{Weygaert1989},
         \citet{pierre1990},
         \citet{barrow1990},
         \citet{coles1991},
         \citet{Weygaert1991a},
         \citet{Weygaert1991b},
         \citet{Subba1992},
         \citet{Ikeuchi1991}
         and
         \citet{Goldwirth1995}.
         An updated review of the 3D Voronoi Diagrams 
         applied to  cosmology can be found in 
          \citet{Weygaert2002} 
          or 
          \citet{Weygaert2003}.
         The 3D Voronoi tessellation was first
        applied to identify groups of galaxies in
         the structure of a super-cluster, 
         see
         \citet{Ebeling1993},
         \citet{Bernardeau1996},
         \citet{Schaap2000},
         \citet{Marinoni2002},
           \citet{Melnyk2006},
         \citet{Schaap2009} 
         and  
         \citet{Elyiv2009}.
The  physical models 
that produce the observed properties of galaxies
are  intimately  related, 
for example through  the Lagrangian approximation, 
and can be 
approximately classified as 
\begin{itemize}
\item {\bf Cosmological N-body}: Through N-body experiments 
       by 
      \citet{Aarseth1978} 
      it  is possible to simulate 
      groups which are   analogous to the studies of groups 
      among bright Zwicky-catalog galaxies, see
      \citet{Gott1979a} 
      or 
       covariance functions in  simulations of galaxy
       clustering in an expanding universe 
       which are  found to be power laws in the nonlinear regime 
       with slopes centered on 1.9
       \citet{Gott1979b}.
       Using gigaparticle N-body simulations to study galaxy 
       cluster populations in Hubble volumes,
       \citet{Evrard2002} 
       created
       mock sky surveys of dark matter structure to z~=1.4 
       over $10000^{\circ}~sq.~deg$ and to
       z~=0.5 over two full spheres.
       In short,
       N-body calculations seek to model
       the full nonlinear system by making discrete the matter 
       distribution 
       and following its evolution in a
       Lagrangian fashion,
       while N-body simulations are usually understood 
       to concern gravity only.

\item {\bf Dynamical Models}: Starting from a power law 
      of primordial inhomogeneities it is possible 
      to obtain a two-point correlation function 
      for galaxies with an exponent similar to that 
      observed, see
      \citet{Peebles1974a,Peebles1974b,Gott1975}.

        Another line of work is to assume that
        the velocity field is of a potential type; 
        this assumption is 
        called the Zel'dovich  approximation, 
         see 
         \citet{ZelDovich1970,ZelDovich1989,Coles1995}.
        The Zel'dovich formalism
        is    a Lagrangian approximation
        of the fully nonlinear set of equations. 
        In this sense it is ``gravity'' only and does not include
        a pressure term.

\item   {\bf The halo models}:
         The halo model
         describes nonlinear structures as virialized dark-matter
         halos of different mass, placing them in space according
         to the linear large-scale density field which is completely
         described by the initial power spectrum,
         see  
         \citet{Neyman1952,Scherrer1991,Cooray2002}.
         Figure~19 in 
         \citet{Jones2005}, for example,
         reports the exact nonlinear model matter distribution   
         compared with its halo-model representation.

\end{itemize}

The absence of clear information on the 3D displacement
of the physical results as a function of the redshift 
and the selected magnitude characterize  
the cosmological N-body,
the dynamical and the hydrodynamical models.
This  absence  of detailed information 
leads to   the analysis of the following questions:
\begin {itemize}
\item Is it possible to
      compare the theoretical and observational 
      number of galaxies  as a  function of the redshift
      for a fixed flux/magnitude ?
\item  What is the role of the Malmquist bias 
      when theoretical and observed numbers  of
      galaxies versus the redshift are compared? 

\item Is it possible to find an algorithm which  describes
      the intersection between a slice 
       that  starts from the center of the box 
       and   
      the faces of  irregular  Poissonian Voronoi Polyhedrons?
\item Is it possible to model the intersection between a sphere
      of a given redshift and the faces of 
       irregular  Poissonian Voronoi Polyhedrons?
\item Does the developed theory  match   the observed 
      slices of galaxies as given, for example,
      by the 
      2dF Galaxy Redshift Survey?
\item Does the developed algorithm explain
      the voids  appearance 
      in all sky surveys such as the RC3?
\item Can  voids between galaxies be modeled 
      trough the Voronoi normalized volume distribution?

\item Is it possible to evaluate the probability 
      of  having  a supervoid once the averaged void's 
      diameter is fixed?

\item Is it possible to compute the correlation 
      function  for galaxies by introducing the 
      concept of  thick faces  
      of  irregular Voronoi  polyhedrons?      

\item Is it possible to find the acoustic oscillations
      of the correlation function at $\approx~100Mpc$ 
      in   simulated  slices 
      of  the Voronoi  diagrams. 

\end{itemize}
In order  to answer these questions,  
Section~\ref{formulary} briefly reviews 
 the standard 
luminosity function for galaxies.
An accurate  test of the number of galaxies as a  function
of the redshift is performed on 
the 2dF Galaxy Redshift Survey (2dFGRS),
see Section~\ref{test}.
Section~\ref{Voronoisec}   reports the technique 
which allows us  to extract  the galaxies belonging to 
the Voronoi polyhedron and  Section~\ref{cellular}   
simulates the redshift dependence of  the  2dFGRS
as well as the overall  
Third Reference Catalog of Bright Galaxies (RC3).
Section  \ref{sec_corr} reports the simulation of the 
correlation function computed on the thick faces of 
the Voronoi polyhedron.

\section{Useful   formulas}
\label{formulary}

Starting from  
\citet{Hubble1929} 
the suggested correlation
between expansion velocity  and distance is
\begin {equation}
V= H_0 D  = c_l \, z  
\quad ,
\label {clz}
\end{equation}
where the Hubble constant  is 
 $H_0 = 100 h \mathrm{\ km\ s}^{-1}\mathrm{\ Mpc}^{-1}$, with $h=1$
when  $h$ is not specified,
$D$ is the distance in $Mpc$,
$c_l$ is  the  light velocity  and
$z$   is  the redshift.
Concerning   the value of  $H_0$ we will adopt
a recent value as obtained by
the 
Cepheid-calibrated luminosity of Type Ia supernovae,
see \citet{Sandage2006},
\begin{equation}
H_0 =(62.3 \pm 5 ) \mathrm{\ km\ s}^{-1}\mathrm{\ Mpc}^{-1}
\quad .
\end {equation}

The quantity $c_lz$, a velocity, or $z$, a number,
characterizes the catalog of galaxies.

We recall that 
the galaxies have peculiar velocities, 
making the measured redshifts a combination of
cosmological redshift plus a contribution 
on behalf of the peculiar velocity.

The  maximum redshift here considered is   $z\approx 0.1 $
meaning a maximum velocity of expansion of $\approx $
30000 $\frac{Km}{s}$;   up 
to that value  the  space is assumed to be Euclidean.
We now report 
the joint distribution in {\it z}  and {\it f} 
(the flux of radiation)  for galaxies 
adopting the Schechter function for
the luminosity ($L$) of galaxies, $\Phi (L)$, introduced by 
\citet{schechter} and 
the mass-luminosity relationship, $\Psi (L)$,  
as derived 
in \citet{Zaninetti2008}.
The joint distribution in
{\it z}  and {\it f} for 
the Schechter function,
see formula~(1.104) in
 \citet{pad} 
or formula~(1.117) 
in 
\citet{Padmanabhan_III_2002} 
,
 is
\begin{equation}
\frac{dN}{d\Omega dz df} =  
4 \pi  \bigl ( \frac {c_l}{H_0} \bigr )^5    z^4 \Phi (\frac{z^2}{z_{crit}^2})
\label{nfunctionz}  
\quad ,
\end {equation}
where $d\Omega$, $dz$ and  $ df $ represent the differential of
the solid angle , the redshift and the flux respectively.
The critical value of $z$,   $z_{crit}$, is 
\begin{equation}
 z_{crit}^2 = \frac {H_0^2  L^* } {4 \pi f c_l^2}
\quad .
\end{equation} 
The number of galaxies, $N_S(z,f_{min},f_{max})$  
comprised between a minimum value of flux,
 $f_{min}$,  and  maximum value of flux $f_{max}$,
can be computed  through  the following integral 
\begin{equation}
N_S (z) = \int_{f_{min}} ^{f_{max}}
4 \pi  \bigl ( \frac {c_l}{H_0} \bigr )^5    z^4 \Phi (\frac{z^2}{z_{crit}^2})
df
\quad .
\label{integrale_schechter} 
\end {equation}
This integral does not  have  an analytical solution 
and therefore 
a numerical integration must be performed.

The number of galaxies in {\it z} and {\it f} as given by 
formula~(\ref{nfunctionz})  has a maximum  at  $z=z_{pos-max}$,
where 
\begin{equation}
 z_{pos-max} = z_{crit}  \sqrt {\alpha +2 }
\label{massimoschecher}
\quad ,
\end{equation} 
where $\alpha$ sets the slope for low values of $L$.
This position   can be re-expressed   as
\begin{equation}
 z_{pos-max} =
\frac
{
\sqrt {2+\alpha}\sqrt {{10}^{ 0.4\,{\it M_{\sun}}- 0.4\,{\it M^*}}}{
\it H_0}
}
{
2\,\sqrt {\pi }\sqrt {f}{\it c_l}
}
\quad  ,
\label{zmax_sch}
\end{equation}
where $M_{\sun}$ is the reference magnitude 
of the sun at the considered bandpass
and $M^*$ is the characteristic magnitude as derived from the
data.
The joint distribution in $z$ and $f$,
in presence of the ${\mathcal M}-L$
relationship, see equation~(38)
\citet{Zaninetti2008},   is 
\begin{equation}
\frac{dN}{d\Omega dz df} =  
4 \pi  \bigl ( \frac {c_l}{H_0} \bigr )^5    z^4 \Psi (\frac{z^2}{z_{crit}^2})
\label{nfunctionz_mia}  
\quad .
\end {equation}
The number of galaxies, 
$N_{{\mathcal M}-L}(z,f_{min},f_{max})$  
with flux  
comprised between 
 $f_{min}$  and   $f_{max}$,
can be computed  through  the following integral 
\begin{equation}
N_{{\mathcal M}-L} (z) = \int_{f_{min}} ^{f_{max}}
4 \pi  \bigl ( \frac {c_l}{H_0} \bigr )^5    z^4 \Psi (\frac{z^2}{z_{crit}^2})
df 
\quad ,
\label{integrale_mia} 
\end {equation}
and also in this case  
a numerical integration must be performed.

The number of galaxies as  given  by 
the ${\mathcal M}-L$ relationship  
 has a maximum at 
$z_{pos-max}$ , see equation~(41) in \citet{Zaninetti2008}
\begin{equation}
 z_{pos-max} = z_{crit} 
\left( {\it c}+a \right) ^{a/2}
\quad ,
\label{zmassimomia}
\end{equation} 
which can be re-expressed as
\begin{equation}
 z_{pos-max} = 
\frac
{
\left( a+{\it c} \right) ^{1/2\,a}\sqrt {{10}^{ 0.4\,{\it M_{\sun}}-
0.4\,{\it M^*}}}{\it H_0}
}
{
2\,\sqrt {\pi }\sqrt {f}{\it c_l}
}
\quad ,
\label{zmax_mia}
\end{equation} 
where  $1/a$ is  an exponent  which  connects
 mass to 
luminosity
and $c$  represents the dimensionality 
of the fragmentation.
 
\section{Photometric test on the catalog}

\label{test}
We now check the  previously derived 
formulas
 on a catalog 
of galaxies.
A first example  is  the 2dFGRS data release 
available on   the  web site: http://msowww.anu.edu.au/2dFGRS/.
In particular we added together the file parent.ngp.txt which  
contains 145652 entries for NGP strip sources and 
the file parent.sgp.txt which 
contains 204490 entries for SGP strip sources.
Once the   heliocentric redshift  was  selected 
we  processed 219107 galaxies with 
$0.001 \leq z \leq 0.3$.
The parameters  of the Schechter function
concerning the 2dFGRS
can be found in the  first line of Table~3 in  
\citet{Madgwick_2002}
and 
are reported in   Table~\ref{parameters}.
It is  interesting to point out  that 
other  values   
for $h$ 
different from  1
shift all absolute magnitudes by $5\log_{10}h$ and change
number densities by the factor $h^3$.

\begin{table}
 \caption[]{The parameters of the Schechter function  for \\
      the 
     2dFGRS as in Madgwick et al. 2002. }
 \label{parameters}
 \[
 \begin{array}{lc}
 \hline
 \hline
 \noalign{\smallskip}
parameter            & 2dFGRS                                  \\ \noalign{\smallskip}
M^* - 5\log_{10}h ~ [mags]         &  ( -19.79 \pm 0.04)           \\ \noalign{\smallskip}
\alpha               &   -1.19  \pm 0.01                       \\ \noalign{\smallskip}
\Phi^* ~[h^3~Mpc^{-3}] &   ((1.59   \pm 0.1)10^{-2})      \\ \noalign{\smallskip}
 \hline
 \hline
 \end{array}
 \]
 \end {table}

 \begin{table} 
 \caption[]{The parameters of the 
             ${\mathcal M}-L$ luminosity function \\
            based on the   2dFGRS data 
           ( triplets  generated by the author). } 
 \label{para_physical} 
 \[ 
 \begin{array}{lc} 
 \hline 
~     &   2dFGRS   \\ \noalign{\smallskip}  
 \hline 
 \noalign{\smallskip} 
c                   &    0.1                 \\ \noalign{\smallskip}
M^*   - 5\log_{10}h [mags]       &  -19  \pm 0.1       \\ \noalign{\smallskip}
\Psi^* [h^3~Mpc^{-3}] &  0.4  \pm 0.01      \\ \noalign{\smallskip}
a                   &  1.3   \pm 0.1       \\ \noalign{\smallskip} 
 \hline 
 \hline 
 \end{array} 
 \] 
 \end {table}

\begin{figure*}
\begin{center}
\includegraphics[width=10cm]{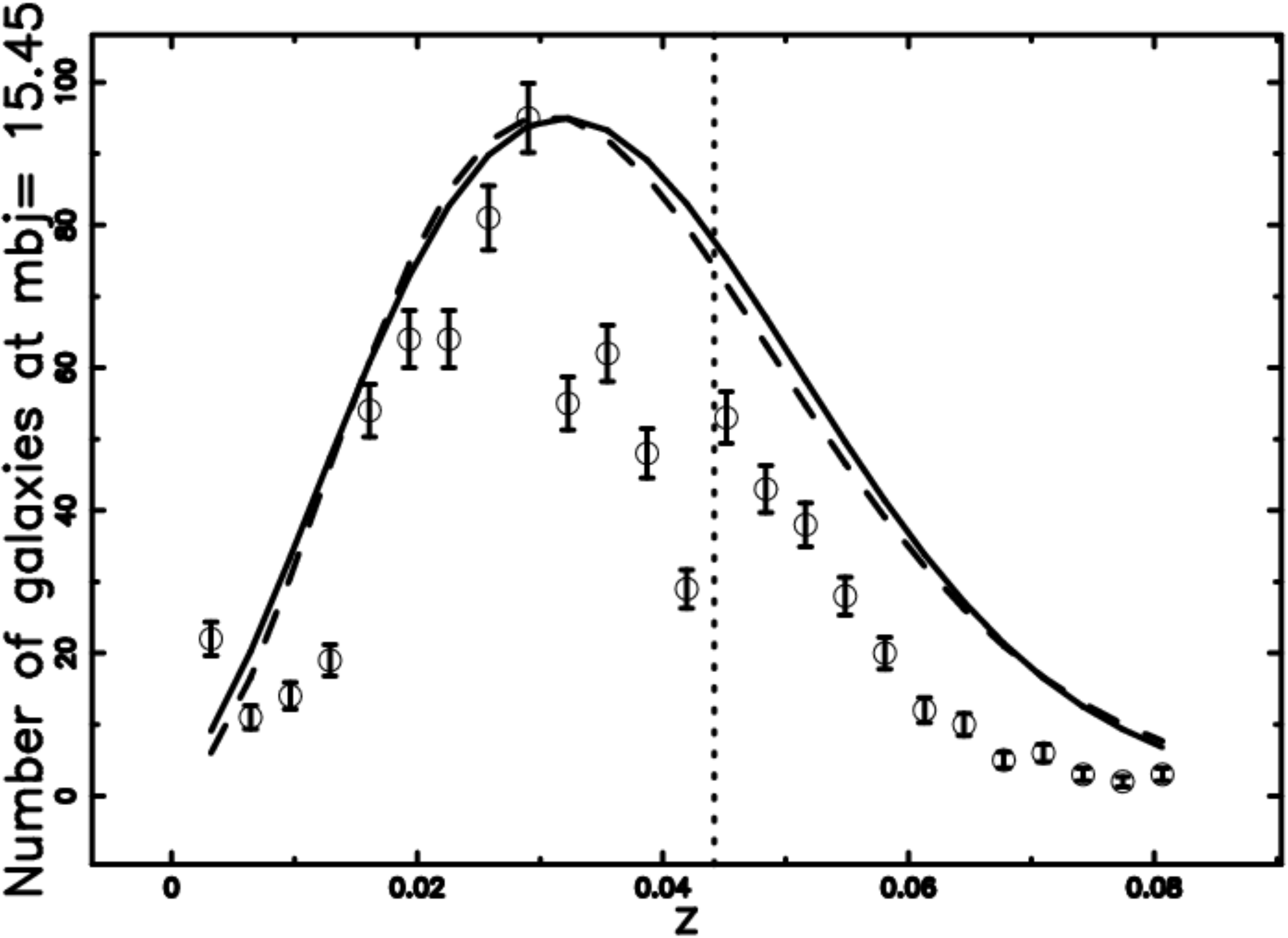}
\end {center}
\caption{
The galaxies  of the 2dFGRS with 
$ 15.27  \leq  bJmag \leq 15.65 $  or 
$ 59253  \frac {L_{\sun}}{Mpc^2} \leq  
f \leq 83868  \frac {L_{\sun}}{Mpc^2}$
( with $bJmag$ representing  the 
relative magnitude  used in object selection),
are isolated 
in order to represent a chosen value of $m$ 
and then organized in frequencies versus
heliocentric  redshift,  
(empty circles);
the error bar is given by the square root of the frequency.
The maximum in the frequencies of observed galaxies is 
at  $z=0.03$.
The maximum of the observed galaxies 
can also be computed through 
the  maximum
likelihood estimator (MLE) by adopting 
 the Schechter function
for
the luminosity, 
see Appendix~A;   
$\widehat  z_{pos-max} =0.033$  according to equation
(\ref{massimoschecherlike}). 
The theoretical curve  generated by
the Schechter  function of luminosity 
(formula~(\ref{nfunctionz}) and parameters
as in column 2dFGRS of Table~\ref{parameters}) 
is drawn  (full line).
The theoretical curve  generated by
the ${\mathcal M}-L$   function for luminosity (
formula~(\ref{nfunctionz_mia})
and  parameters as in column  2dFGRS of Table~\ref{para_physical})
is drawn  (dashed line);
 $\chi^2$= 550  for the Schechter  function and $\chi^2$= 503
for the ${\mathcal M}-L$   function.
In this plot $\mathcal{M_{\sun}}$ = 5.33  and $h$=0.623.
The vertical  dotted line represents the boundary
between complete and incomplete samples.
}
          \label{maximum_flux}%
    \end{figure*}

\begin{figure*}
\begin{center}
\includegraphics[width=10cm]{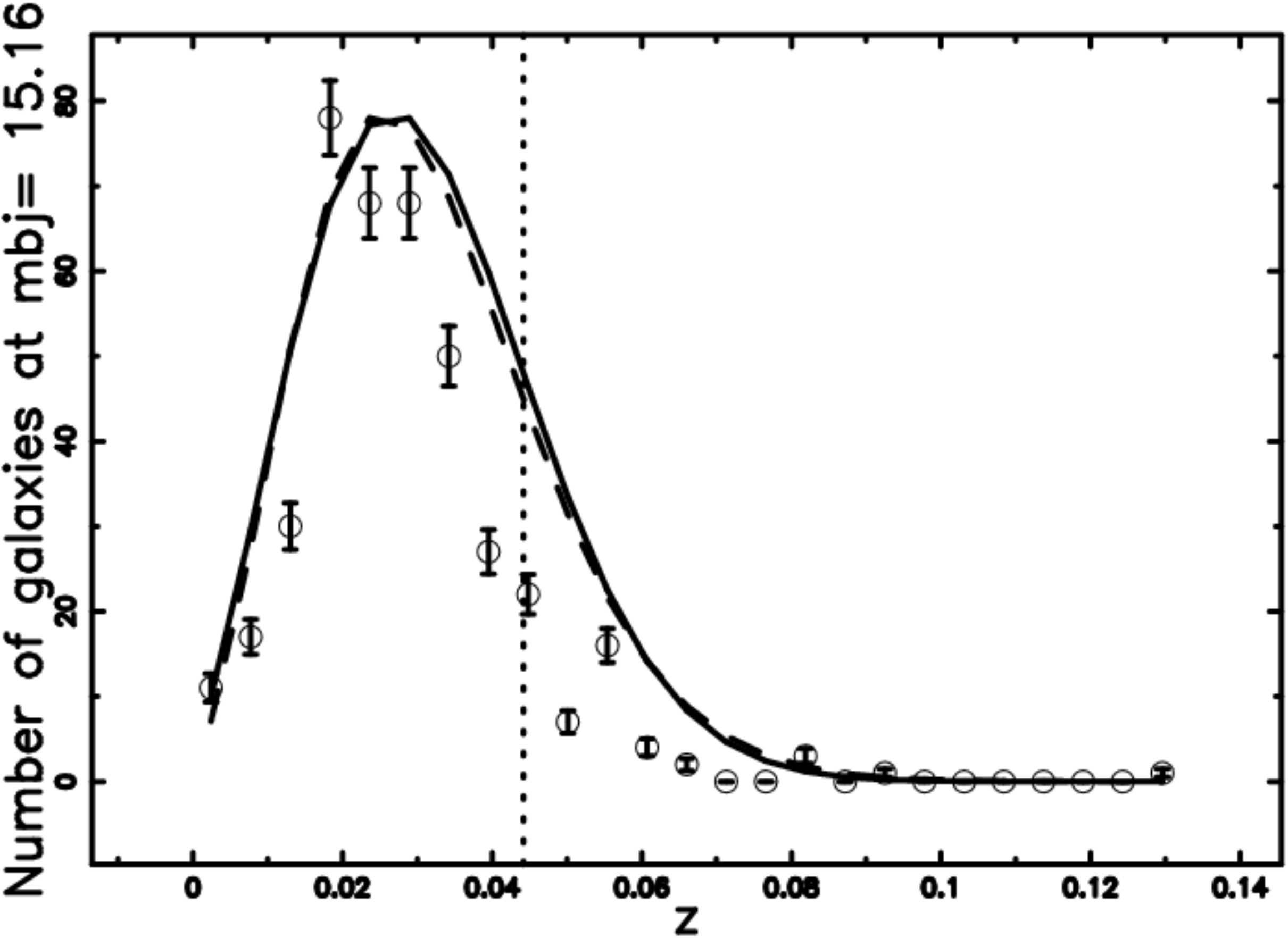}
\end {center}
\caption{
The galaxies  in the  2dFGRS  with 
$ 15.02 \leq  bJmag \leq 15.31 $
or 
$ 80527  \frac {L_{\sun}}{Mpc^2} \leq  
f \leq 105142 \frac {L_{\sun}}{Mpc^2}$.
The maximum in the frequencies of observed galaxies 
is at  $z=0.02$,
 $\chi^2$= 256  for the Schechter  function (full line)
 and $\chi^2$= 224
for the ${\mathcal M}-L$   function (dashed line).
The maximum of the observed galaxies 
can also be computed through 
the  maximum
likelihood estimator (MLE) by adopting the 
 the Schechter function
for
the luminosity, 
see Appendix~A;   
$\widehat  z_{pos-max} =0.031$  according to equation
(\ref{massimoschecherlike}). 
In this plot $\mathcal{M_{\sun}}$ = 5.33  and $h$=0.623.
The vertical  dotted line represents the boundary
between complete and incomplete samples.
}
          \label{maximum_flux_2}%
    \end{figure*}

Before  reducing  the data we should discuss the Malmquist bias,
see \citet{Malmquist_1920,Malmquist_1922}, 
that was originally applied
to the stars and was then 
 applied to the galaxies by \citet{Behr1951}. 
We
therefore introduce
 the concept of
limiting apparent magnitude and the correspondent
 completeness in
absolute magnitude of the considered catalog as a function of the
redshift. The observable absolute magnitude as a 
function of the
limiting apparent magnitude, $m_L$, is
\begin{equation}
M_L =
m_{{L}}-5\,{\it \log_{10}} \left( {\frac {{\it c_L}\,z}{H_{{0}}}}
 \right) -25
\quad .
\label{absolutel}
\end{equation}
The previous formula predicts,  from a theoretical
point of view the upper limit of the absolute
maximum magnitude that can be observed in a
catalog of galaxies characterized by a given limiting
magnitude.
The interval covered by the LF for galaxies, 
$\Delta M $,
is defined as
\begin{equation}
\Delta M = M_{max} - M_{min}
\quad ,
\end{equation}
where $M_{max}$ and $M_{min}$ are the
maximum and minimum
absolute
magnitude of the LF for the considered catalog.
The real observable interval in absolute magnitude,
$\Delta M_L $,
 is
\begin{equation}
\Delta M_L = M_{L} - M_{min}
\quad .
\end{equation}
We can therefore introduce the range
of observable absolute maximum magnitude
expressed in percent, $ \epsilon(z) $,
as
\begin{equation}
\epsilon_s(z) = \frac { \Delta M_L } {\Delta M } \times 100
\quad .
\label{range}
\end{equation}
This is a number that represents the completeness of the sample
and given the fact that the limiting magnitude of the 2dFGRS is
$m_L$=19.61 it is possible to conclude that the 2dFGRS is complete
for $z\leq0.0442$.
Figure~\ref{maximum_flux}
and Figure~\ref{maximum_flux_2}
report the number of  observed  galaxies
of the 2dFGRS  catalog for  two different  
apparent magnitudes  and
two theoretical curves  as represented by 
formula~(\ref{nfunctionz})   and  formula~(\ref{nfunctionz_mia}).

\begin{figure*}
\begin{center}
\includegraphics[width=10cm]{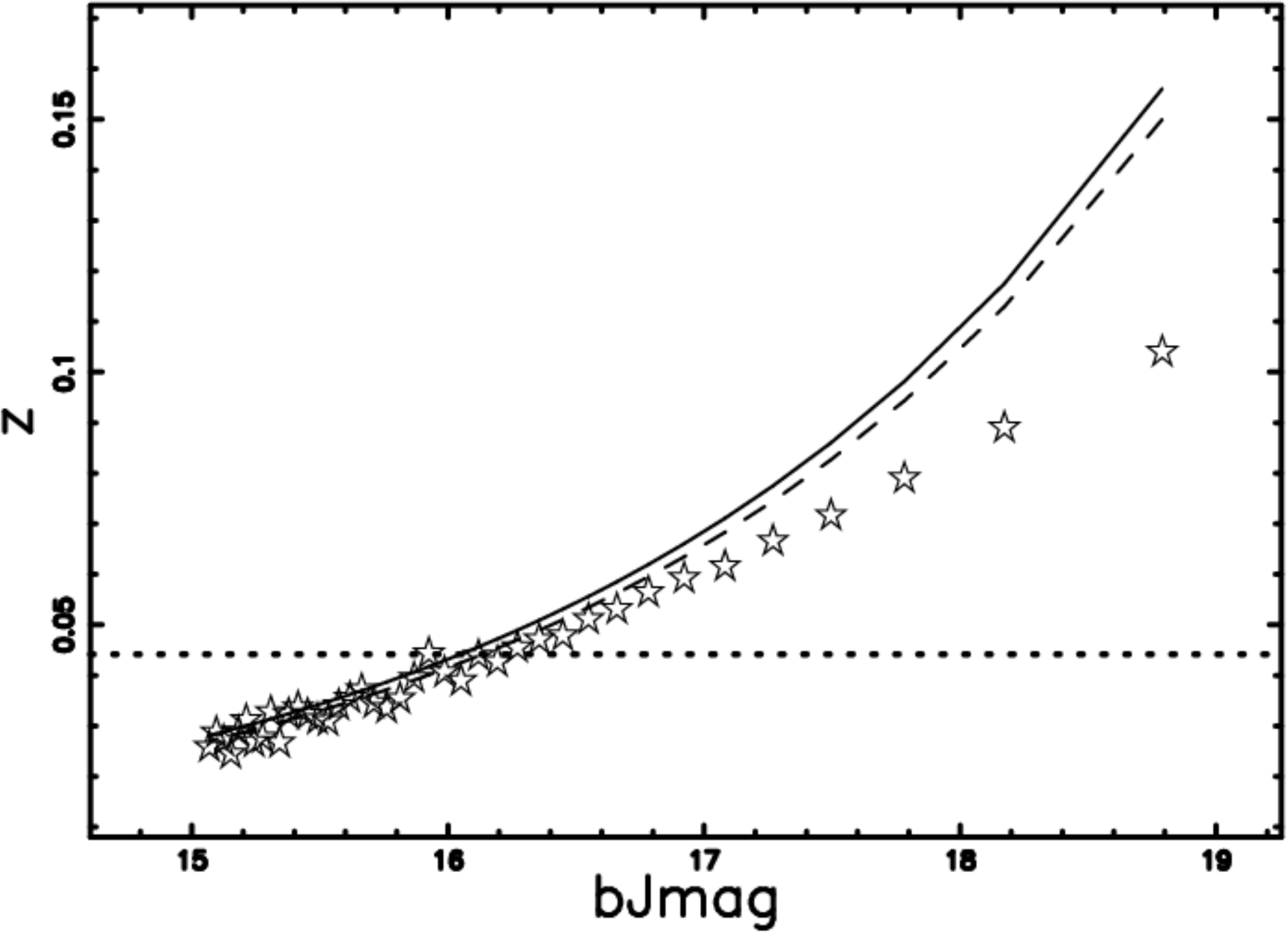}
\end {center}
\caption{
Value of  
$\widehat  z_{pos-max}$ 
( see equation~\ref{massimoschecherlike}) 
 at which  the number of
 galaxies  in the 2dFGRS 
is maximum as a function of 
the   apparent magnitude $ bJmag$ 
(stars),
theoretical curve of the maximum for the 
Schechter  function  as represented by  formula~(\ref{zmax_sch}) 
(full line)   and
theoretical curve of the maximum for the 
${\mathcal M}-L$   function as represented by  formula~(\ref{zmax_mia}) 
(dashed line).
In this plot $\mathcal{M_{\sun}}$ = 5.33  and $h$=0.623.
The horizontal   dotted line represents the boundary
between complete and incomplete samples.
}
          \label{zeta_max_flux}%
    \end{figure*}

Due to the importance of the maximum as a function of  $z$ in the number of 
galaxies,   Figure~\ref{zeta_max_flux}  reports 
the observed histograms in the 2dFGRS
and the theoretical curves  as a  function of the  magnitude.

Following is an
 outline of the sources of discrepancy between theory 
(equations~(\ref{nfunctionz})  and  (\ref{nfunctionz_mia}))
\begin{itemize}
\item The density of galaxies is assumed to be constant
      in deriving the theoretical equations.
      In a  cellular structure of the universe with 
      the galaxies situated on the faces of 
      irregular polyhedrons, the number of 
      galaxies varies with $r^2$, where $r$ 
      is the progressive
      distance from the center of the box, up
      to a distance equal to the averaged diameter of
      a polyhedron.
      After that distance the number of galaxies
      grows as $r^3$.
\item The interval in magnitude should be chosen in order
      to be smaller than the error in magnitude.
\item A limited range in $z$ should be considered in 
      order to satisfy the Malmquist bias.
\end{itemize}

\begin{figure*}
\begin{center}
\includegraphics[width=10cm]{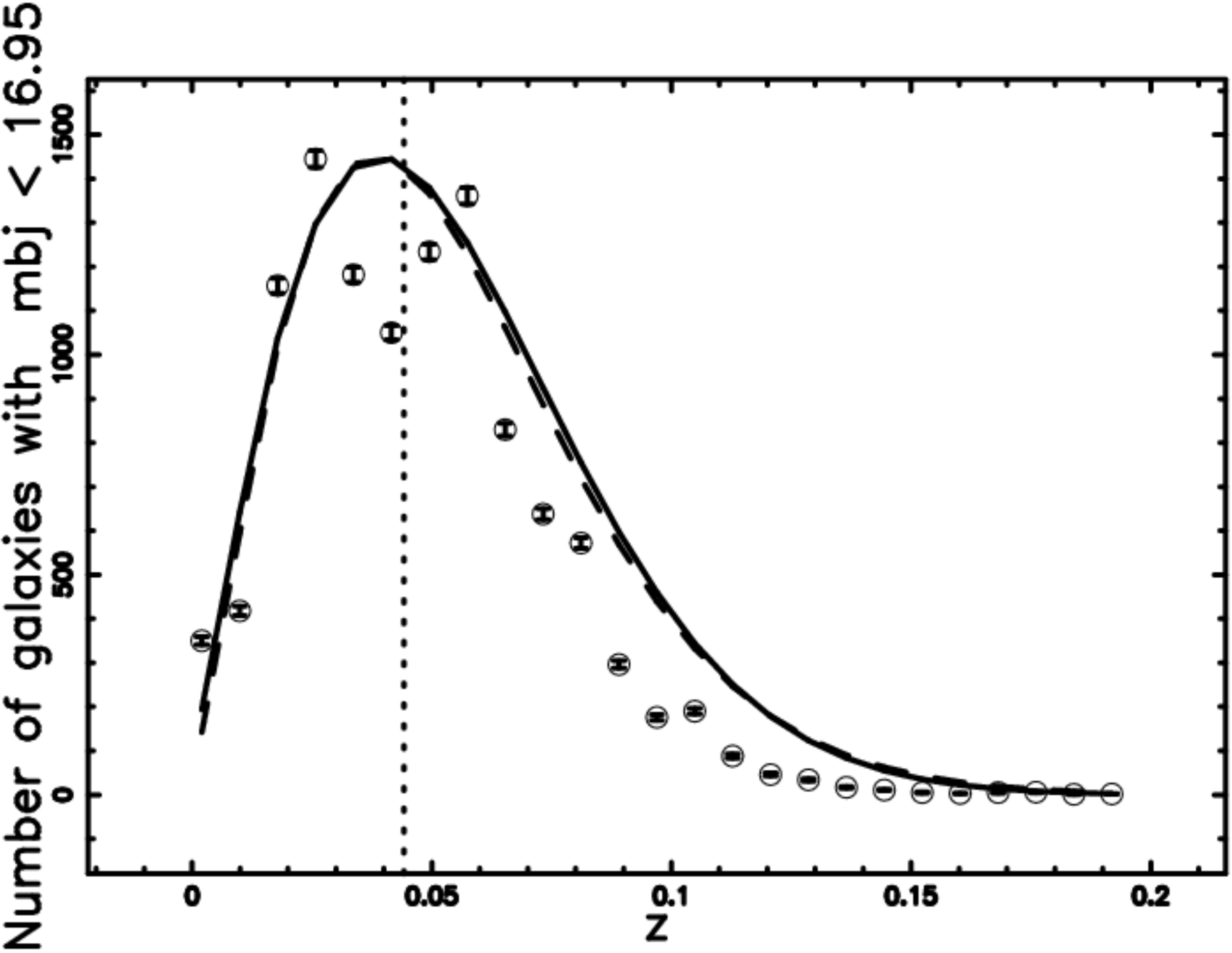}
\end {center}
\caption{The galaxies in the  2dFGRS     with 
$ 13.34 \leq  bJmag \leq 16.94 $
or 
$ 17950  \frac {L_{\sun}}{Mpc^2} \leq  
f \leq 493844 \frac {L_{\sun}}{Mpc^2}$,
are organized in  frequencies versus
heliocentric  redshift,  (empty stars).
The theoretical curves  generated by
the integral of the Schechter  function  in flux  
(formula~(\ref{integrale_schechter})  with parameters
as  in  Table~\ref{parameters}) 
(full  line)
and by the integral of the  
${\mathcal M}-L$   function as represented by  
formula~(\ref{integrale_mia}) 
with   parameters as in column  2dFGRS of Table~\ref{para_physical})
(dashed line)  are drawn.
The maximum in the frequencies of observed galaxies 
is at  $z=0.029$,
 $\chi^2$= 3314  for the Schechter  function (full line)
 and $\chi^2$= 3506
for the ${\mathcal M}-L$   function (dashed line).
In this plot $\mathcal{M_{\sun}}$ = 5.33  and $h$=0.623.
The vertical  dotted line represents the boundary
between complete and incomplete sample.
}
          \label{maximum_flux_all}%
    \end{figure*}

The total number of galaxies in the 2dFGRS
is reported in Figure~\ref{maximum_flux_all}  as well 
as the theoretical curves as represented 
by the numerical integration of formula~(\ref{nfunctionz})
and formula~(\ref{nfunctionz_mia}).
In this section we have adopted the absolute magnitude of
the sun in the $b_j$ filter 
$\mathcal{M_{\sun}}$ = 5.33, see
\citet{Einasto_2009,Eke_2004}.

\section{The 3D Voronoi Diagrams  }

The observational fact  that  the galaxies 
seem to be distributed on
almost bubble like
surfaces, 
surrounding  large empty regions 
allows us to introduce the geometrical properties of 
irregular Voronoi Polyhedron   
as a useful tool to explain the  galaxy's network.
The faces of the Voronoi Polyhedra     
share the same property , i.e.
they are equally distant from 
two nuclei.  The intersection 
between a plane and the faces produces diagrams which are similar
to the edges'  displacement in  2D Voronoi diagrams.
From the point of view of the observations
it is very useful to study the intersection 
between a slice which crosses  the center of the box 
and the faces of  irregular polyhedrons where presumably
the galaxies reside.
The general definition of the  3D Voronoi Diagrams
is given in Section~\ref{general}.
The intersection between a slice of a given opening angle, 
for example $3^{\circ}$, 
and the faces of the Voronoi Polyhedra     
can be realized through an approximate  algorithm,
see  next  Section~\ref{faces}.
The volumes of the Voronoi Polyhedra can be identified 
by  the voids between galaxies, while  the statistics 
that describe  the volumes 
can help to study  the statistics  
of  the void's  size distribution in  the 2dFGRS,
see 
\citet{Muller2008}.
\label{Voronoisec} 

\subsection{General Definition}

\label{general}
The Voronoi diagram for a set of  seeds, $S$,
located at position $x_i$  
in $\mathcal{R}^3$ space  is the
partitioning of that space into regions such that 
all locations within any
one region are closer to the generating point 
than to any other.
The points closer to one seed than another are 
divided by the perpendicular
bisecting plane
between the two seeds. 
For
a random tessellation, only a finite number of half-planes 
bind  the cell,
so the cell is a convex
polyhedron. 
It follows that a plane cross-section of a 3D
tessellation is a tessellation of
the plane composed of convex polygons.
The points of a  3D tessellation  are 
of four types, depending
on how many nearest
neighbors in $S$ they have. 
A point with exactly one nearest neighbor is in
the interior of a cell, a point
with two nearest neighbors is on the 
face between two cells, 
a point with
three nearest neighbors is
on an edge shared by three cells, 
and a point with four neighbors is a
vertex where three cells meet.
There is zero probability 
that there will be 
any point with five or more
nearest neighbors.
In the following we will work on a three dimensional
lattice defined by 
$ pixels \times pixels \times pixels $ 
points,
$L_{kmn}$.

The Voronoi polyhedron $V_i$ around
a given center $i$ , 
is the set of lattice points $L_{kmn}$  closer
to $i$ than to any $j$: more formally,
\begin{equation}
  L_{kmn} \; \epsilon \;   V_i \leftrightarrow \mid x_{kmn} - x_i \mid 
\leq \mid x_{kmn} - x_j
\quad  ,  
\end{equation}
where $ x_{kmn}$ denotes the lattice point position. 
Thus,  the Polyhedra  are intersections
of half-spaces. Given a center $i$ and its neighbor $j$,  
the line
$ij$ is cut perpendicularly at its midpoint $ y_{ij}$ by
the plane $h_{ij}$. 
$H_{ij}$ is the half-space generated 
by the plane $h_{ij}$, which  consists of 
the subset of lattice points on 
the same side of $ h_{ij}$ as  $i$ ; therefore 
\begin{equation}
     V_i = \cap _j H_{ij }    ,    
\end{equation}
$V_i$ is bounded by faces , 
with each face $ f_{ij} $ belonging 
to a distinct plane $h_{ij}$. 
Each face will be characterized 
by its vertices and edges.

\subsection{The adopted algorithm} 

\label{faces}
Our  method considers a  3D lattice with 
${\it pixels}^{3}$ points:
present in this lattice are $N_s$ seeds generated
according to a random process.
All the computations are usually performed on this mathematical
lattice; 
the conversion to the physical lattice
is obtained   by multiplying the unit
by $\delta=\frac{side}{pixels -1}$, where {\it side}
is the  length of the cube expressed in the physical
unit  adopted.
In order to minimize boundary  effects 
introduced by those
polyhedron which  cross the cubic boundary,
the cube in which the seeds are inserted 
is amplified
by  a factor {\it amplify}. 
Therefore the $N_s$  seeds are inserted in a volume
   $  pixels^3 \times
amplify$,  
which is bigger than  the box over which  the
scanning is performed; 
{\it amplify } is generally taken to be equal to 1.2.
This procedure inserts periodic boundary conditions to our
cube. 
A sensible and solid discussion of what such an 
extension of a cube should be 
can be  found  in 
\citet{Neyrinck2005b}.
The set $S$ of the seeds can be of
Poissonian   or non-Poissonian type.
Adopting  the point of view that the universe should be the same
from each point of view of the observer the 
Poissonian seeds can represent the best choice
in order to reproduce the large scale structures.

The Poissonian   seeds  are generated independently on the $X$, 
$Y$  and
$Z$ axis in 3D through a subroutine  which returns a
pseudo-random real number taken from a uniform distribution
between 0 and 1. For practical purposes,
 the subroutine
RAN2  was used, see \citet{press}.
Particular attention  should be paid to
the average observed diameter of voids, $\overline{DV^{obs}}$,
here chosen 
as
\begin{equation}
\overline{DV^{obs}} \approx  0.6  {DV_{max}^{obs}} = 2700 \frac{Km}{sec} 
\quad ,
\label{dvobserved}
\end {equation} 
where $DV_{max}^{obs}=4500~\frac{Km}{sec}$  corresponds to the 
extension of the maximum void visible, 
for example,   on the CFA2 slices, see \citet{geller}.
The corresponding diameter, $\overline{DV^{obs}}$,
in $pc$  is
\begin{equation}
\overline{DV^{obs}} = \frac{27}{h}~Mpc 
\quad .
\label{dobserved}
\end {equation}

The number of Poissonian     seeds is chosen in such 
a way that the averaged 
volume occupied by a Voronoi polyhedron is equal to 
the averaged observed volume of  the voids 
in the spatial distribution
of galaxies; 
more details can be found in
\citet{Zaninetti2006}.
It is possible to plot the 
cumulative volume-weighted void size distribution, $F(>R)$, 
in the 2dFGRS samples,
see Figure 4 in 
\citet{Muller2008}.
From the previous figure it is possible  to make a graphical
evaluation of the value of $R$
at which  $F(>R)=1/2$, the median  of the 
probability density function   connected with 
$F(>R)$.
The median value of $R$ from Figure 4 in 
 \citet{Muller2008}  
turns out to be 
$5~Mpc<R< 12~Mpc$ according to the 
four models there implemented.
The average value of $R$ here 
assumed to be $\frac{13.5}{h}~Mpc$ is not far 
from the median value presented in  
\citet{Muller2008}.

\begin{figure*}
\begin{center}
\includegraphics[width=10cm]{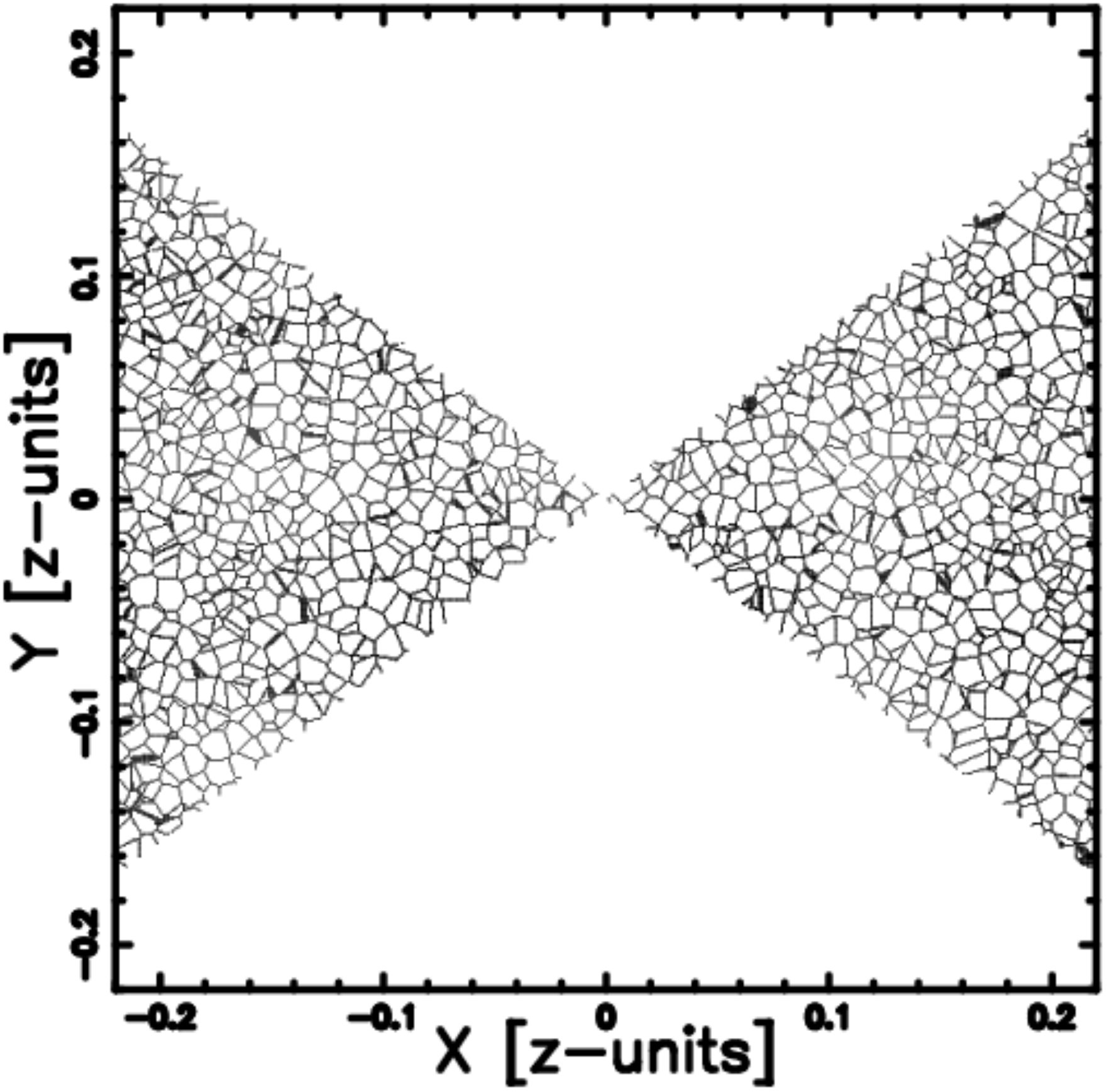}
\end {center}
\caption{
Portion of the  Poissonian Voronoi--diagram $V_p(2,3)$ ;
cut on the  X-Y plane   when two strips of 
$75^{\circ}$ are considered.
The  parameters
are      $ pixels$= 600
       , $ N_s   $   = 137998
       , $ side  $   = 131908 $Km/sec$
  and    $ amplify$~= 1.2~.}
          \label{cut_middle}%
    \end{figure*}

We now work on a 3D lattice  L$_{k,m,n}$ of $pixels^3$
 elements.
Given a section  of the cube
(characterized, for example, by $k=\frac{pixels}{2}$)
the  various $V_i$ (the volume belonging
to the seed i)
 may or may not cross the  pixels 
 belonging to the two dimensional lattice.
A typical  example of   a 2D cut  organized in   two  strips 
about $75^{\circ}$ long  
is  visible  in Figure~\ref{cut_middle} 
where 
the Cartesian coordinates $X$ and $Y$ with the origin 
of the axis at the center of the box  has been used.
The previous cut has an extension on the 
$Z$-axis equal to zero.

\begin{figure*}
\begin{center}
\includegraphics[width=10cm]{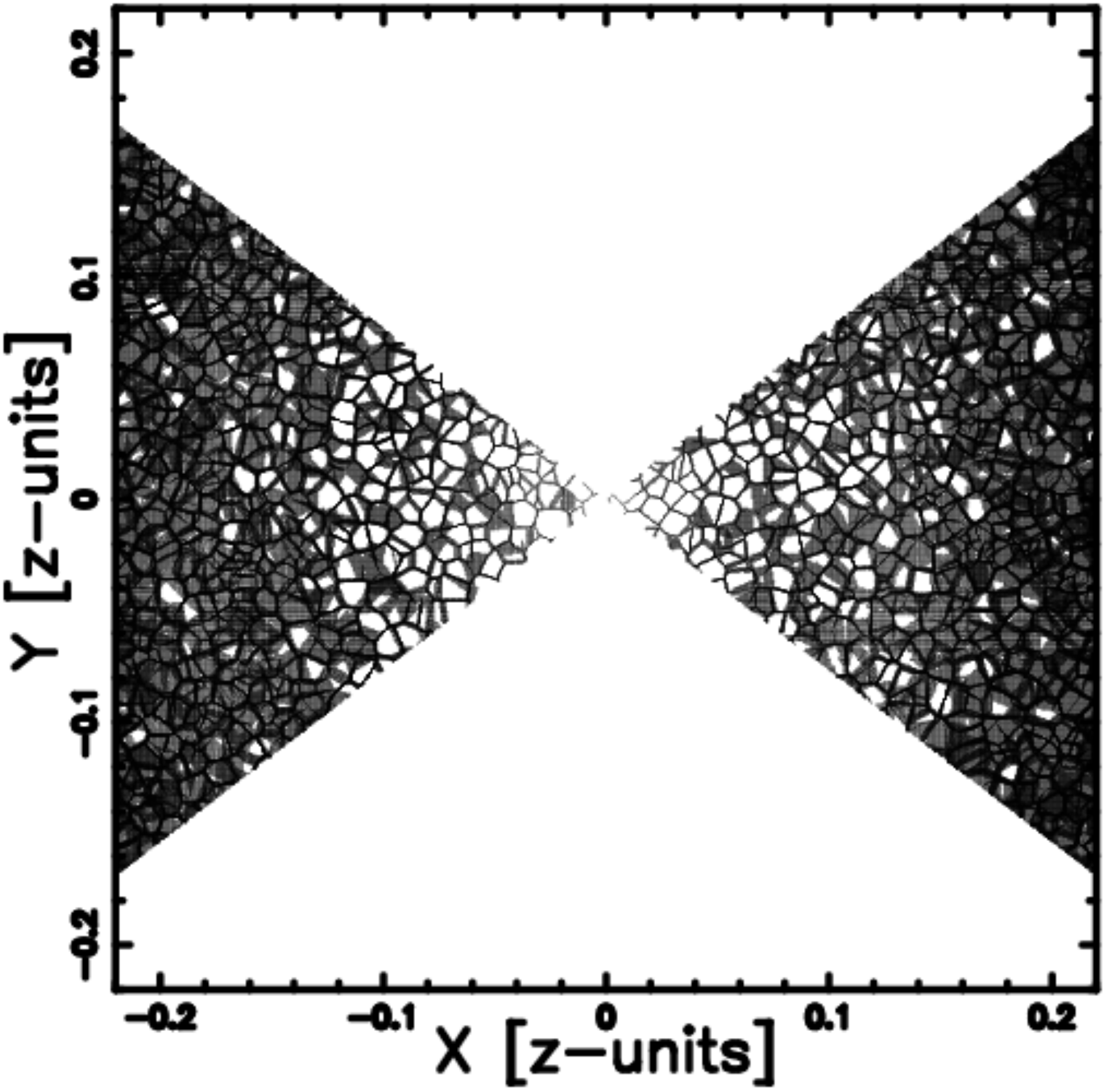}
\end {center}
\caption{
The same as Figure~\ref{cut_middle}
but now   two slices  of 
$75^{\circ}$  long and  $3^{\circ}$
wide are considered.} 
          \label{voro_fetta_tutte}%
    \end{figure*}

Conversely Figure~\ref{voro_fetta_tutte} reports 
two slices
of $75^{\circ}$  long and  $3^{\circ}$
wide. 
In this case the extension of the enclosed region 
belonging to the $Z$-axis increases with distance 
according to
\begin{equation}
\Delta Z = \sqrt {X^2 +y^2} \tan \frac {\alpha} {2} 
\quad ,
\end{equation}
where $\Delta Z $ is the thickness of the slice and
$\alpha$ is  the opening angle, 
in our case $3^\circ$.

In order to simulate the slices of observed galaxies 
 a subset is extracted ( randomly chosen)   
of the pixels belonging
to a slice  as represented, for example,
in  Figure~\ref{voro_fetta_tutte}. 
In this operation of extraction of the galaxies from the pixels
of the slice, the photometric rules as represented by 
formula~(\ref{nfunctionz}) must  be respected.   

\begin{figure}
\begin{center}
\includegraphics[width=10cm]{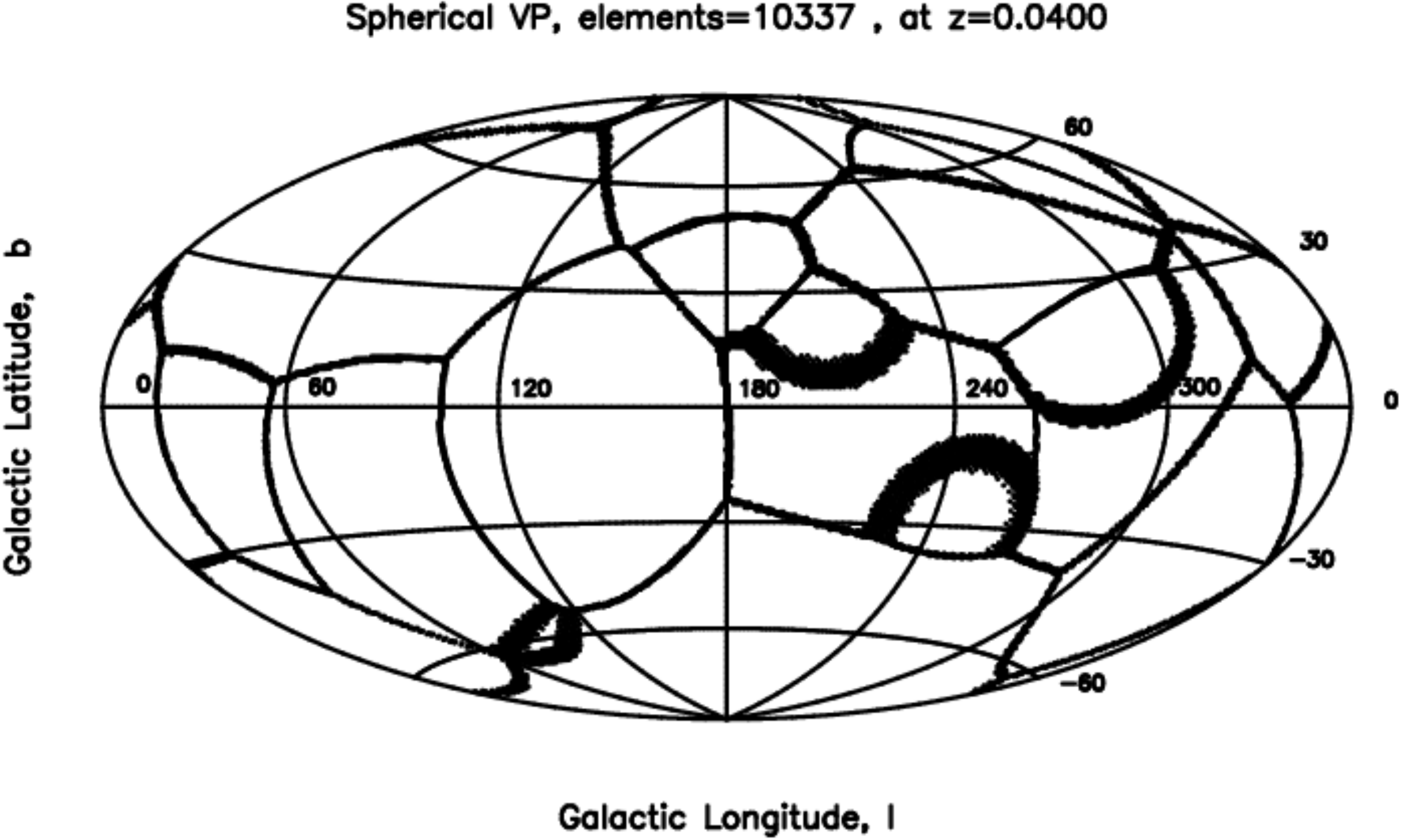} 
\end {center}
\caption {
The Voronoi--diagram $V_s(2,3)$ 
in the Hammer-Aitoff  projection
at $z$ = 0.04.
The  parameters
are      $ pixels$= 400, 
         $ N_s   $   = 137998, 
         $ side  $   = 131908 $Km/sec$
and    $ amplify$= 1.2.}
          \label{aitof_sphere}%
    \end{figure}

\begin{figure}
\begin{center}
\includegraphics[width=10cm]{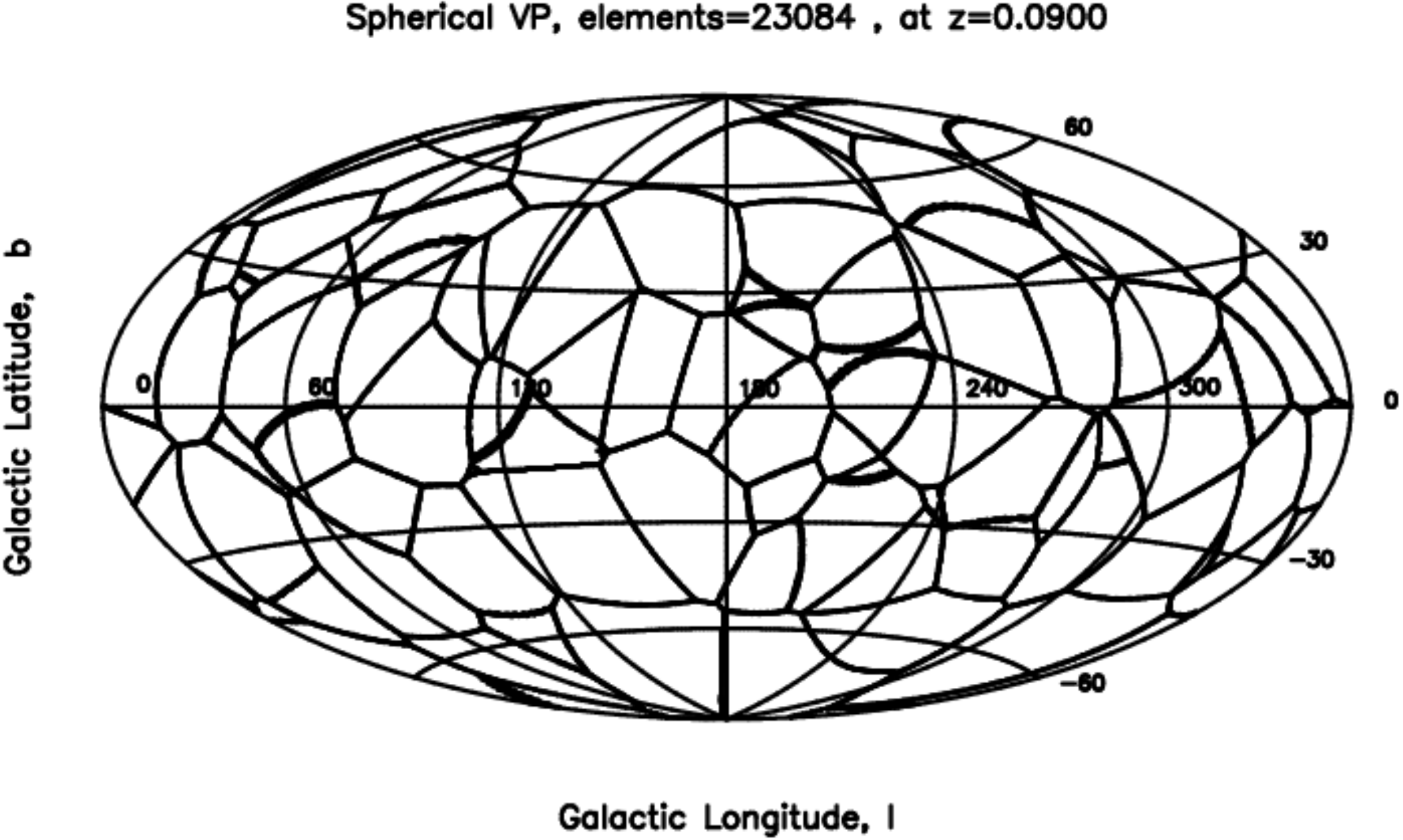} 
\end {center}
\caption {
The Voronoi--diagram $V_s(2,3)$ 
in the Hammer-Aitoff  projection
at $z$ = 0.09;
other  parameters as in Figure~\ref{aitof_sphere}.
        }
          \label{aitof_sphere_2}%
    \end{figure}

The cross sectional area of the VP can also
be visualized through 
a spherical cut characterized by a constant value 
of the distance to the center of the box,
in this case expressed in $z$ units, 
see Figure~\ref{aitof_sphere}
and Figure~\ref{aitof_sphere_2};
 this intersection is called $V_s(2,3)$  where the 
index $s$ stands for sphere.

\subsection{The statistics of the volumes}

\label{volumes}
\begin{figure}
\begin{center}
\includegraphics[width=10cm]{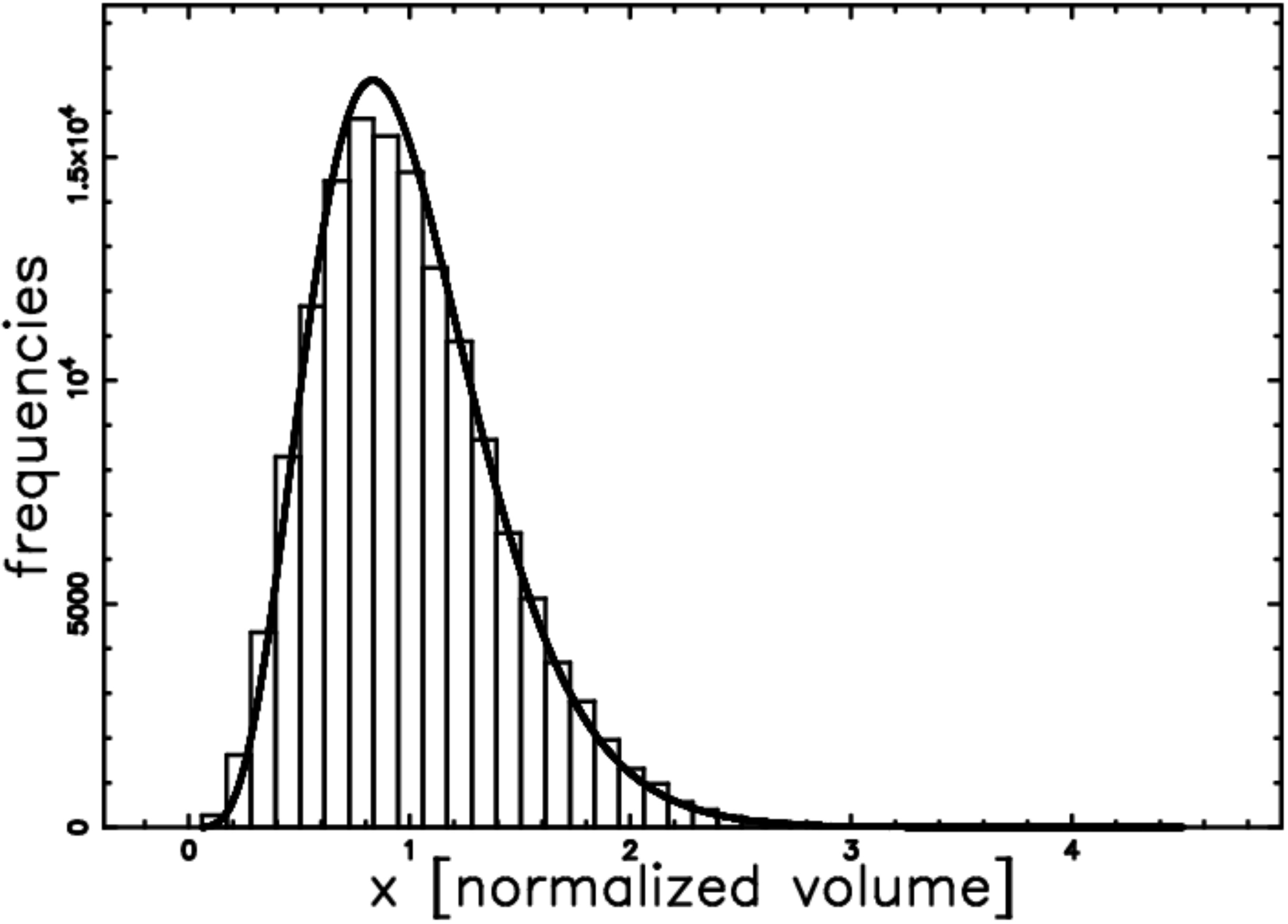} 
\end {center}
\caption {
Histogram (step-diagram)  of 
the Voronoi  normalized volume distribution in 3D
with a superposition of the
gamma--variate as represented by equation~(\ref{kiangd}).
Parameters  as in Figure~\ref{cut_middle}
but  {\it pixels}~= 500: d=3, NBIN=40 and $\chi^2$=1778}
          \label{gamma_kiang_6}%
    \end{figure}

\begin{figure}
\begin{center}
\includegraphics[width=10cm]{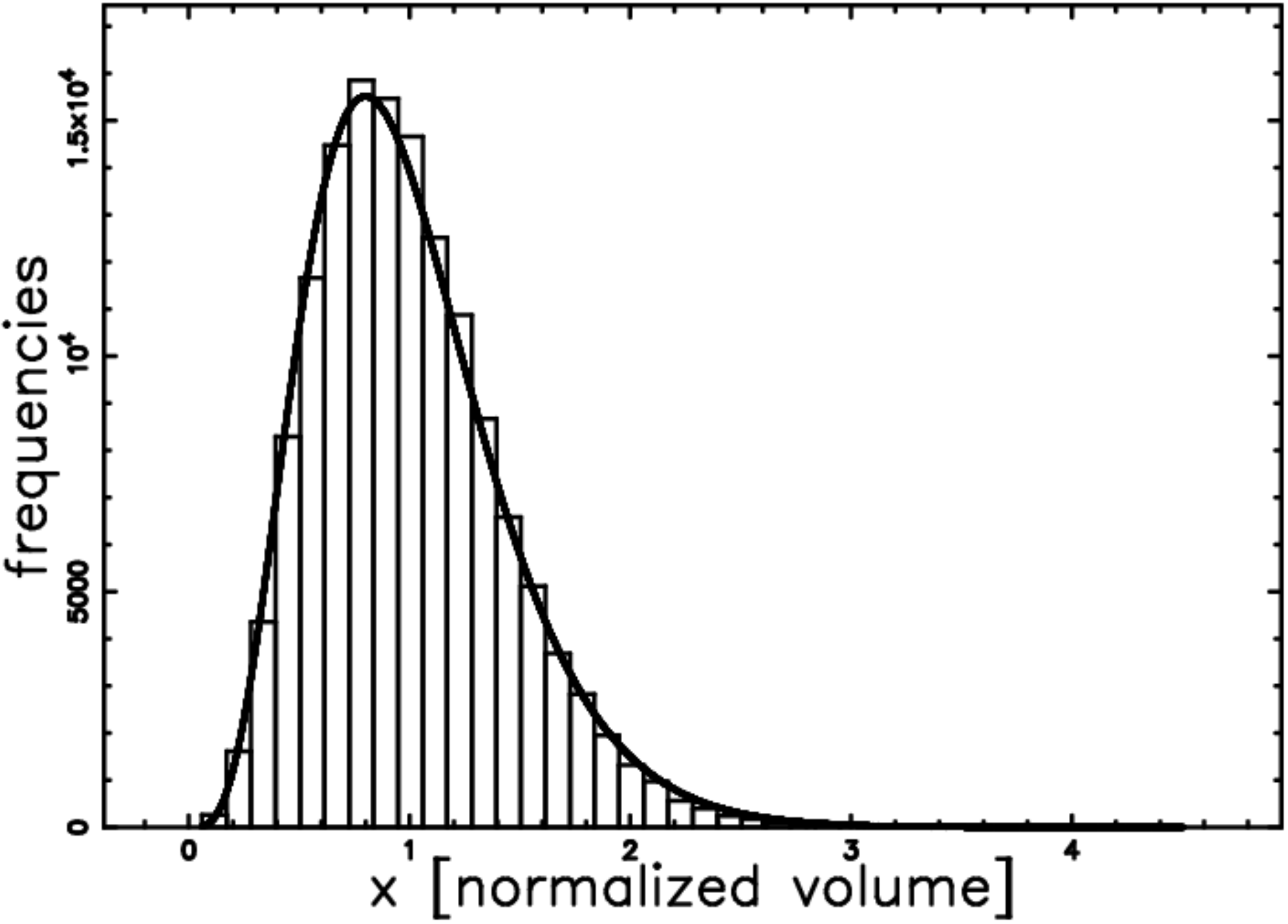} 
\end {center}
\caption 
{
Histogram (step-diagram)  of 
the Voronoi  normalized volume distribution in 3D
with a superposition of the
gamma PDF  as represented by equation~(\ref{rumeni})
first introduced in  
Ferenc \& Neda 2007.
Parameters  as in Figure~\ref{cut_middle}
but   {\it pixels} = 500: d=3, NBIN=40 and $\chi^2$=565
}
\label{gamma_rumeni}%
    \end{figure}

The distribution of volumes in the 
Poissonian Voronoi Diagrams can be modeled
by the following  probability density function
(PDF) $H (x ;c_k )$
\begin{equation}
 H (x ;c_k ) = \frac {c_k} {\Gamma (c_k)} (c_kx )^{c_k-1} \exp(-c_kx)
\quad ,
\label{kiang}
\end{equation}
where $  0 \leq x < \infty $ , $ c_k~>0$
and  $\Gamma (c_k)$ is the gamma function with argument $c_k$,
see formula~(5) by  
\citet{kiang}  .
According to the  "Kiang conjecture"
the volumes should be characterized by $c_k=6$,
see 
\citet{Zaninetti2008} 
for more details.
This PDF  can be  generalized by 
introducing  the 
dimension of the considered space,  $d(d=1,2,3)$,
and  $c_k=2d$  
\begin{equation}
 H (x ;d  ) = \frac {2d} {\Gamma (2d)} (2dx )^{2d-1} \exp(-2dx)
\quad .
\label{kiangd}
\end{equation}
 A new analytical PDF  is 
\begin{equation}
FN(x;d) =  Const \times x^{\frac {3d-1}{2} } \exp{(-(3d+1)x/2)}
\quad ,
\label{rumeni}
\end{equation}
where
\begin{equation}
Const =  
\frac 
{
\sqrt {2}\sqrt {3\,d+1}
}
{
2\,{2}^{3/2\,d} \left( 3\,d+1 \right) ^{-3/2\,d}\Gamma  \left( 3/2\,d+
1/2 \right) 
}
\quad ,
\end{equation}
and $d(d=1,2,3)$ represents the 
dimension of the considered space,  see 
\citet{Ferenc_2007}.
In  the two PDF  here presented, equation~(\ref{rumeni}) and  
equation~(\ref{kiangd}), the statistics of the volumes are
obtained by inserting $d=3$.
In order to obtain the volumes 
in  every  point--lattice L$_{k,m,n}$ 
we computed the nearest 
seed  and 
 increased by one the volume
of that seed.
The  frequency histogram and the relative best fit
through gamma-variate  PDF for the volume distribution
is reported in Figure~\ref{gamma_kiang_6};
Figure~\ref{gamma_rumeni}  conversely reports the fit 
with PDF~(\ref{rumeni}) by   
\citet{Ferenc_2007}.

The results are reported  in Table~\ref{data}
 \begin{table}
 \caption[]{The $\chi^2$  of data fit when the number
            of classes is 40  for three  PDF }
 \label{data}
 \[
 \begin{array}{lc}
 \hline
PDF    &   \chi^2  \\ \noalign{\smallskip}
 \hline
 \noalign{\smallskip}
H (x ;d )  ~when~ d=3  ~ (c_k=6)       &    1778                  \\ \noalign{\smallskip}
H (x ;d )  ~when~ d=2.75~ (c_k=5.5)    &    414               \\ \noalign{\smallskip}
FN(x;d  )~ Ferenc ~\&~ Neda~ formula~(\ref{rumeni}) ~when~ d=3   &    565                  \\ \noalign{\smallskip}
 \hline
 \hline
 \end{array}
 \]
 \end {table}
and it is possible to conclude
that the volume distribution of  irregular     
Voronoi polyhedron
is better described by
PDF~(\ref{rumeni}) in     
\citet{Ferenc_2007}
rather than  the sum of three  gamma variates
with argument 2.
When conversely   $d$ is used 
as a free parameter to be deduced from the sample
in PDF~(\ref{kiangd}) we obtain a smaller 
$\chi^2$ with respect to the function in 
\citet{Ferenc_2007}.

\begin{figure}
\begin{center}
\includegraphics[width=10cm]{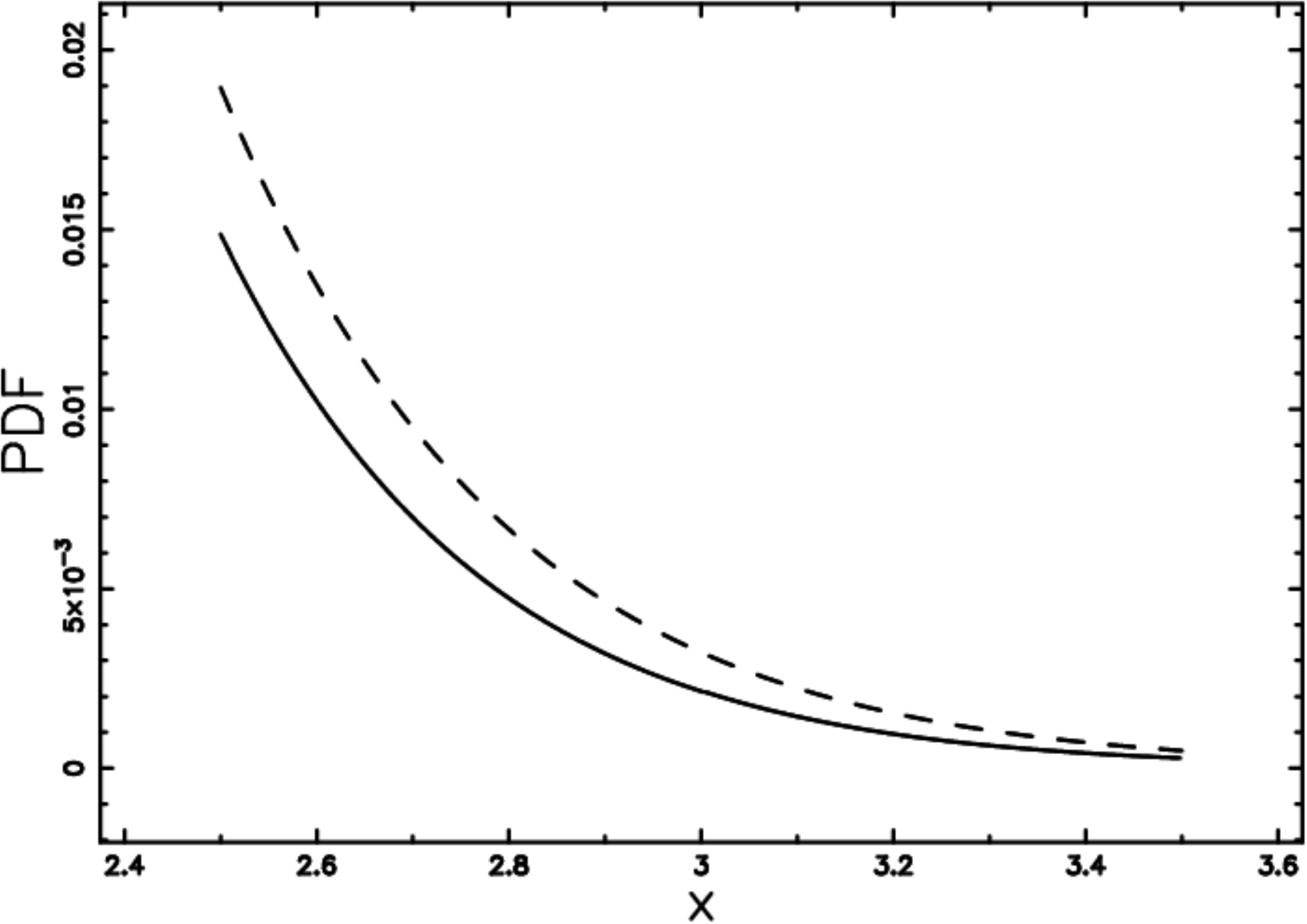}
\end {center}
\caption
{ 
Plot of 
Voronoi  normalized  volume-distribution in 3D 
when 2 PDFs are adopted:
k(x;d)  ~(Eq. (\ref{kiangd})), d=2.75
 (full line)  
and  
FN(x;d) ~(Eq.(\ref{rumeni})), d=3,
(dashed).
}
          \label{confronto}
    \end{figure}

On summarizing the differences  between the Kiang 
function and  the 
Ferenc \& Neda function we can say that the Kiang
function (equation(\ref{kiangd})) requires
the numerical evaluation 
of the free parameter $d=c_k/2$.
In the case of  Ferenc \& Neda (equation~(\ref{rumeni}))
the number of free parameters is zero
once $d=3$ is inserted.
The numerical difference between the two PDFs is 
reported in Figure~\ref{confronto}  for large values of 
the normalized volume distribution.

\section{The cellular structure of the Universe }
\label{cellular}

From a simplified point of view the galaxies belonging 
to a given catalog are characterized by the  astronomical
coordinates, 
the redshift and the apparent magnitude.
Starting from the  second CFA2 redshift   Survey,
the catalogs were organized in slices of a given 
opening angle, $3^{\circ}$ or $6^{\circ}$,
and a given  angular
extension, for example $130^{\circ}$.
When plotted in polar coordinates of $c_lz$ 
the spatial distribution of galaxies
is not random but distributed on filaments.
Particular attention should be paid to the fact that 
the astronomical slices are not a plane which intersects 
a Voronoi Network. 
In order to quantify this effect we introduce a confusion
 distance, $DV_c$, as the distance  after which 
the half altitude of the slices equalizes the 
observed average diameter  $\overline{DV^{obs}}$
\begin{equation}
DV_c \tan (\alpha) = \frac{1}{2} \overline{DV^{obs}}
\quad  ,
\end{equation}
where $\alpha$ is the opening angle  of the slice 
and  $\overline{DV^{obs}}$ the averaged diameter of
voids.
In the case of  2dFGRS   $\alpha=3^{\circ}$ 
and therefore 
$DV_c=2.57~10^{4} \frac {Km}{sec} $
when   $\overline{DV^{obs}}=2700\frac {Km}{sec} $.
For values of $c_lz$ greater than $DV_c$ 
the voids in the distribution
of galaxies are dominated by the confusion.
For values of $c_lz$ lower  than $DV_c$
the filaments of galaxies  can be considered  
the intersection
between a plane and the faces of the Voronoi Polyhedrons.
A measure of the portion of the sky covered by 
a catalog of galaxies is the area covered 
by a  unitarian sphere  which  is 
$4\pi$ steradians or $\frac{129600}{\pi}$ square degrees.
In  the case of 2dFGRS  the covered area 
of two slices  of $75^{\circ}$  long and  $3^{\circ}$
wide,
as in Figure~\ref{2df_cone},
is  $\frac{1414}{\pi}$ square degrees or 0.13 $sr$.
In the case of RC3 the covered area 
it is $4\pi$ steradians with the 
exclusion of the {\it Zone of Avoidance}, see 
Figure~\ref{rc3_all}.
In the following we will simulate 
the 2dFGRS, a catalog that occupies a small area of the sky 
and RC3, a catalog that occupies all the sky.

In the case of 3C3 we  demonstrate  how it is possible to 
simulate the {\it Zone of Avoidance}
in the theoretical simulation.
The paragraph ends with a discussion 
on the Eridanus supervoid also known as "Cold Spot".

\subsection{The 2dFGRS }

\label {cat2dFGRS}

The survey consists of two separate declination strips:
one strip
(the SGP strip) is in the southern Galactic hemisphere 
and covers
approximately $80 ^ {\circ}\times  15^{\circ}$   
centered close 
to the South Galactic Pole.
The second  strip
(the NGP strip) is in the northern Galactic hemisphere 
and covers
$80^{\circ}\times 15^{\circ}$, see \citet{Colless2001}.

\begin{figure*}
\begin{center}
\includegraphics[width=10cm]{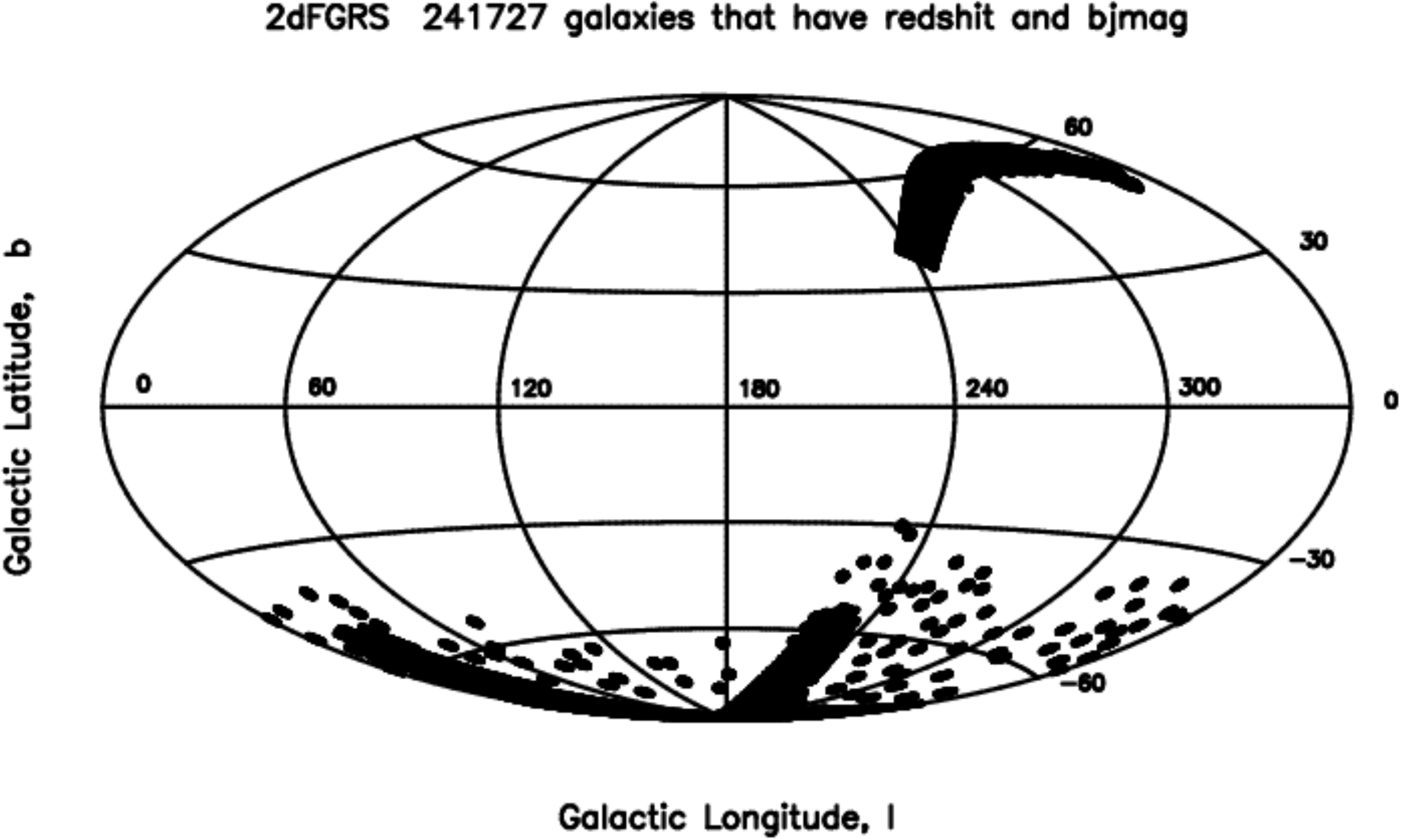}
\end {center}
\caption{Hammer-Aitoff  projection  in galactic coordinates 
  of 230540 galaxies  
 in the 2dfGRS   which have bJmag  and redshift $<0.3$.
}
          \label{2dftuttegal}%
    \end{figure*}

\begin{figure*}
\begin{center}
\includegraphics[width=10cm]{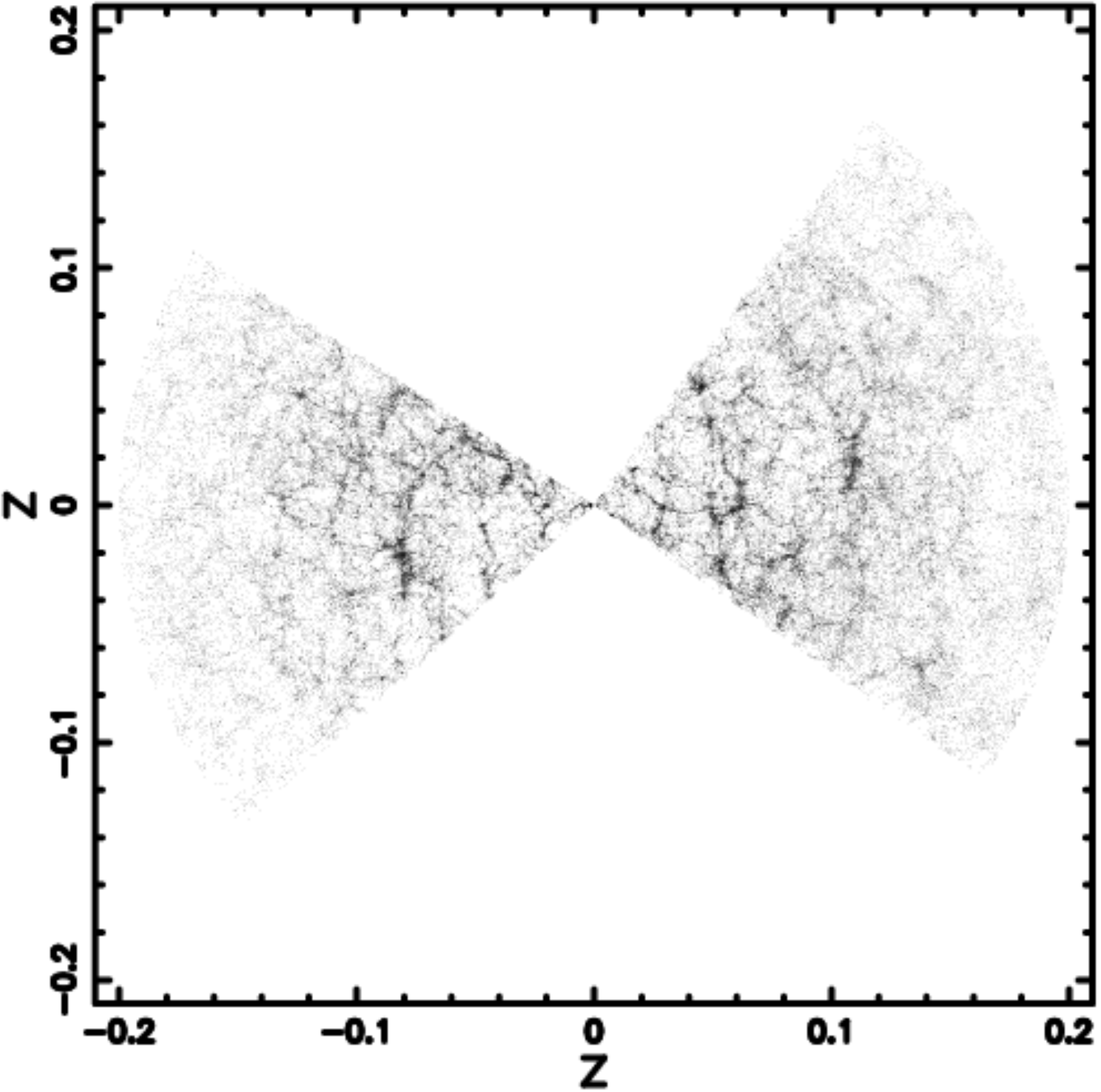}
\end {center}
\caption{Cone-diagram  of all the  galaxies  
in  the 2dFGRS.
This plot contains  203249  galaxies.  
}
          \label{2df_all}%
    \end{figure*}

Figure~\ref{2dftuttegal} shows the galaxies 
of  the 2dFGRS with $z<0.3$ in galactic coordinates
and the  two strips in the  2dFGRS are 
shown in Figure~\ref{2df_all}.

\begin{figure*}
\begin{center}
\includegraphics[width=10cm]{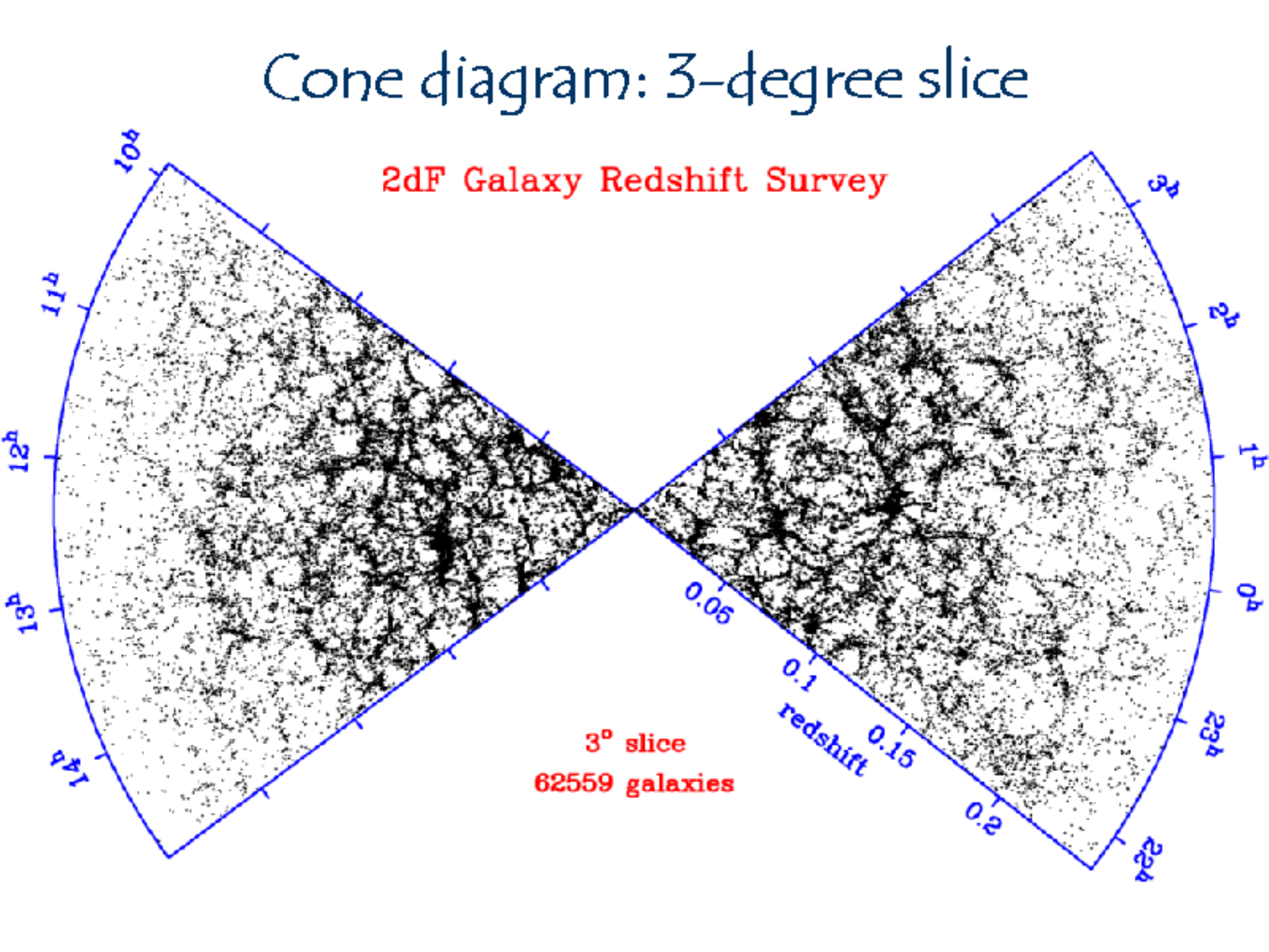}
\end {center}
\caption{Slice  of   $75^{\circ} \times 3^{\circ}$  
in the 2dFGRS.
This plot contains  62559  galaxies and belongs to the   
 2dFGRS Image Gallery
available at  the  web site: http://msowww.anu.edu.au/2dFGRS/.
}
          \label{2df_cone}%
    \end{figure*}

Figure~\ref{2df_cone}  conversely reports  
the 2dfGRS catalog   when a slice of
$75^{\circ} \times 3^{\circ}$
is taken into account.
This slice represents 
the object to simulate.

The previous observational slice can  be simulated
by adopting the Voronoi  network reported 
in Figure~\ref{voro_fetta_tutte}.

\begin{figure*}
\begin{center}
\includegraphics[width=10cm]{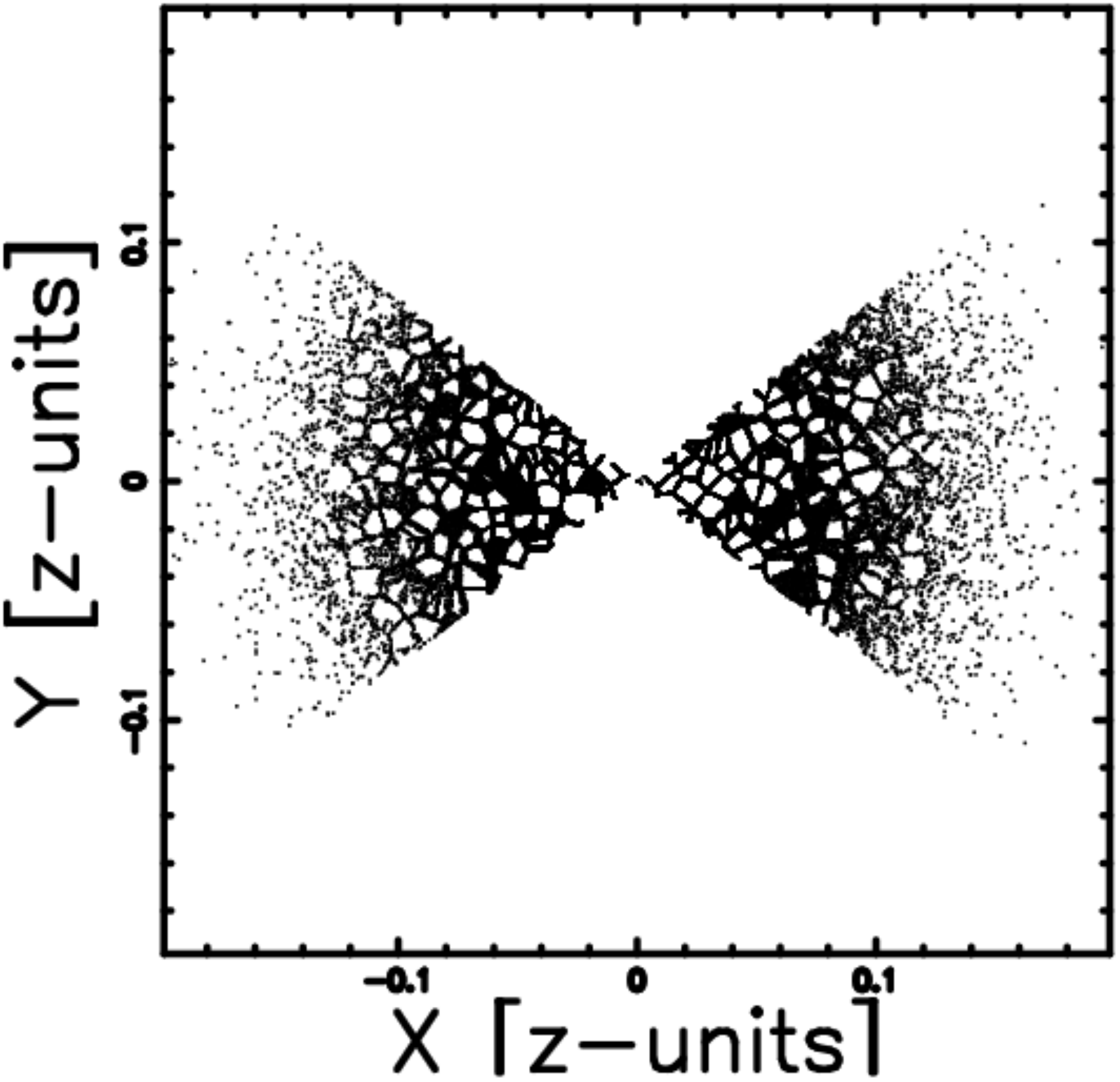}
\end {center}
\caption{
Polar plot
of the  pixels  belonging to a
slice   $75^{\circ}$~long  and $3^{\circ}$
wide.
This plot contains  62563
galaxies,
the maximum in the frequencies of theoretical
galaxies is at  $z=0.043$.
In this plot $\mathcal{M_{\sun}}$ = 5.33  and $h$=0.623.
}
          \label{voro_2df_cones}%
    \end{figure*}

The  distribution of the galaxies as  
given by the Voronoi Diagrams
is reported in Figure~\ref{voro_2df_cones} 
where  all the galaxies  are considered.
In this case the  galaxies are extracted according 
to the integral of the Schechter  function  in flux  
(formula~(\ref{integrale_schechter})  with parameters
as  in  Table~\ref{parameters}). 
Table~\ref{zvalori} reports the basic data of the
astronomical and  simulated data of the 
$75^{\circ} \times 3^{\circ}$ slice.

\begin{table}
\caption { 
Real and  Simulated  data 
of the slice   $75^{\circ}$~long  and $3^{\circ}$.
     }
 \label{zvalori}
 \[
 \begin{array}{ccc}
 \hline
 \hline
 \noalign{\smallskip}
~~   &   2dFGRS   & simulation~         \\
 \noalign{\smallskip}
 \hline
 \noalign{\smallskip}
elements    &  62559  &  62563  \\
z_{min}     &  0.001  &  0.011  \\
z_{pos-max} &  0.029  &  0.042  \\
z_{ave}     &  0.051  &  0.058  \\
z_{max}     &  0.2    &  0.2    \\
\noalign{\smallskip}
\noalign{\smallskip}
 \hline
 \hline
 \end{array}
 \]
 \end {table}

\begin{figure*}
\begin{center}
\includegraphics[width=10cm]{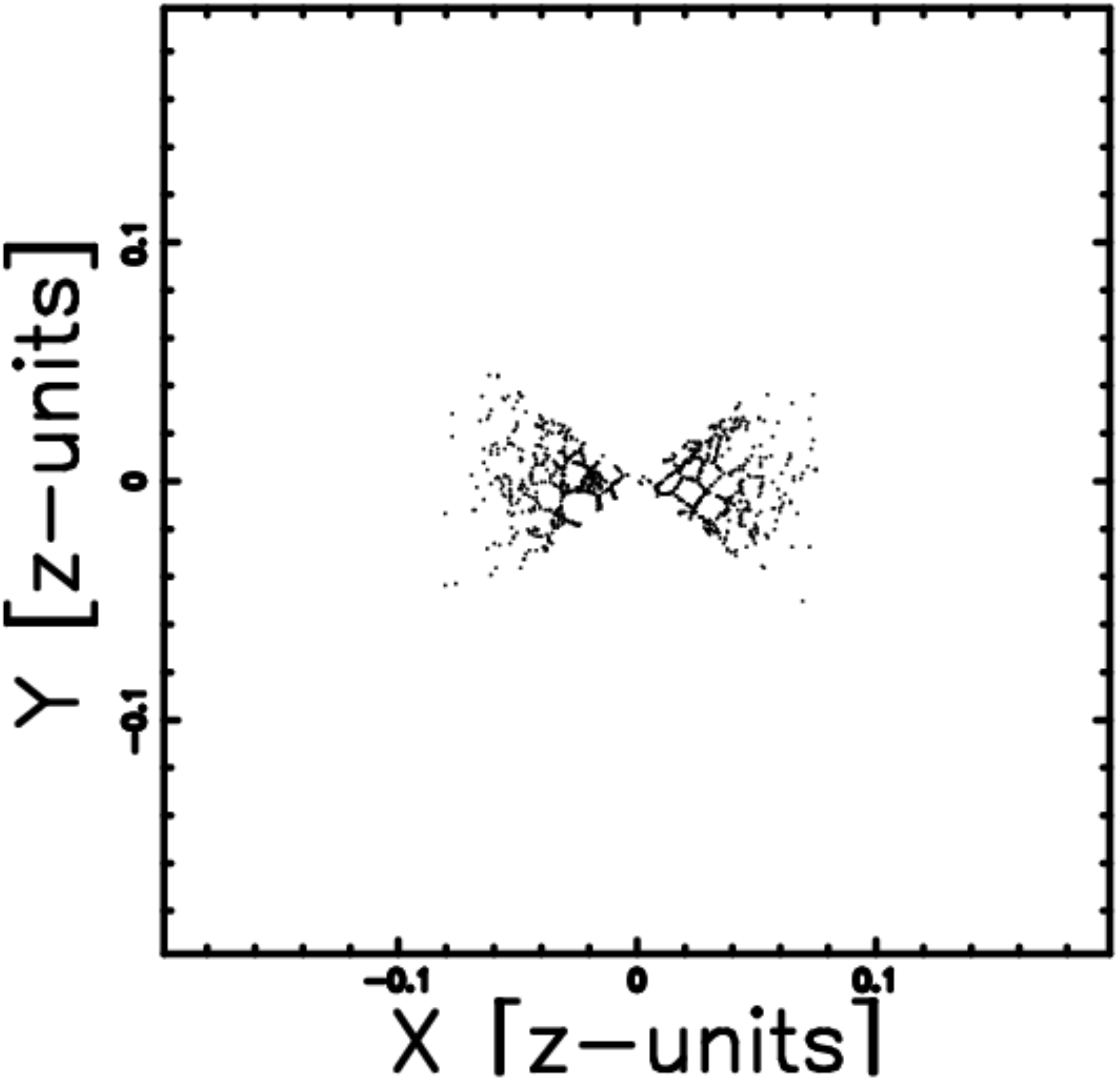}
\end {center}
\caption{
Polar plot
of the  pixels  belonging to a
slice   $75^{\circ}$~long  and $3^{\circ}$
wide.
Galaxies  with magnitude 
$ 15.02  \leq  bJmag \leq 15.31 $
or 
$ 46767  \frac {L_{\sun}}{Mpc^2} \leq  
  61063 \frac {L_{\sun}}{Mpc^2}$.
The maximum in the frequencies of theoretical
galaxies is at  $z=0.029$, 
$N_{SC}$=2186  and    $N_C$=62559.
In this plot $\mathcal{M_{\sun}}$ = 5.33  and $h$=0.623.
}
          \label{voro_2df_cones_sel}%
    \end{figure*}

When conversely a given interval in flux  (magnitudes) 
characterized by  $f_{min}$ and  $f_{max}$
is considered the number of galaxies, $N_{SC}$,
of a $3^{\circ}$ slice
can be found 
with the following formula 
\begin{equation}
N_{SC}  = N_C
\frac  
{ 
\int_{f_{min}} ^{f_{max}}
4 \pi  \bigl ( \frac {c_l}{H_0} \bigr )^5    z^4 \Phi (\frac{z^2}{z_{crit}^2})
df} 
{ 
\int_{f_{min,C}} ^{f_{max,C}}
4 \pi  \bigl ( \frac {c_l}{H_0} \bigr )^5    z^4 \Phi (\frac{z^2}{z_{crit}^2})
df} 
\quad ,
\end{equation}
where   $f_{min,C}$ 
and     $f_{max,C}$  represent
the minimum and maximum flux of the considered catalog
and  $N_C$ all the galaxies of the considered catalog;
a typical  example is reported in 
Figure~\ref{voro_2df_cones_sel}.

\subsection{The Third Reference Catalog of Bright Galaxies}

The RC3, see  \citet{RC31991}, 
is available  at the following address 
http://vizier.u-strasbg.fr/viz-bin/VizieR?-source=VII/155.

\begin{figure*}
\begin{center}
\includegraphics[width=10cm]{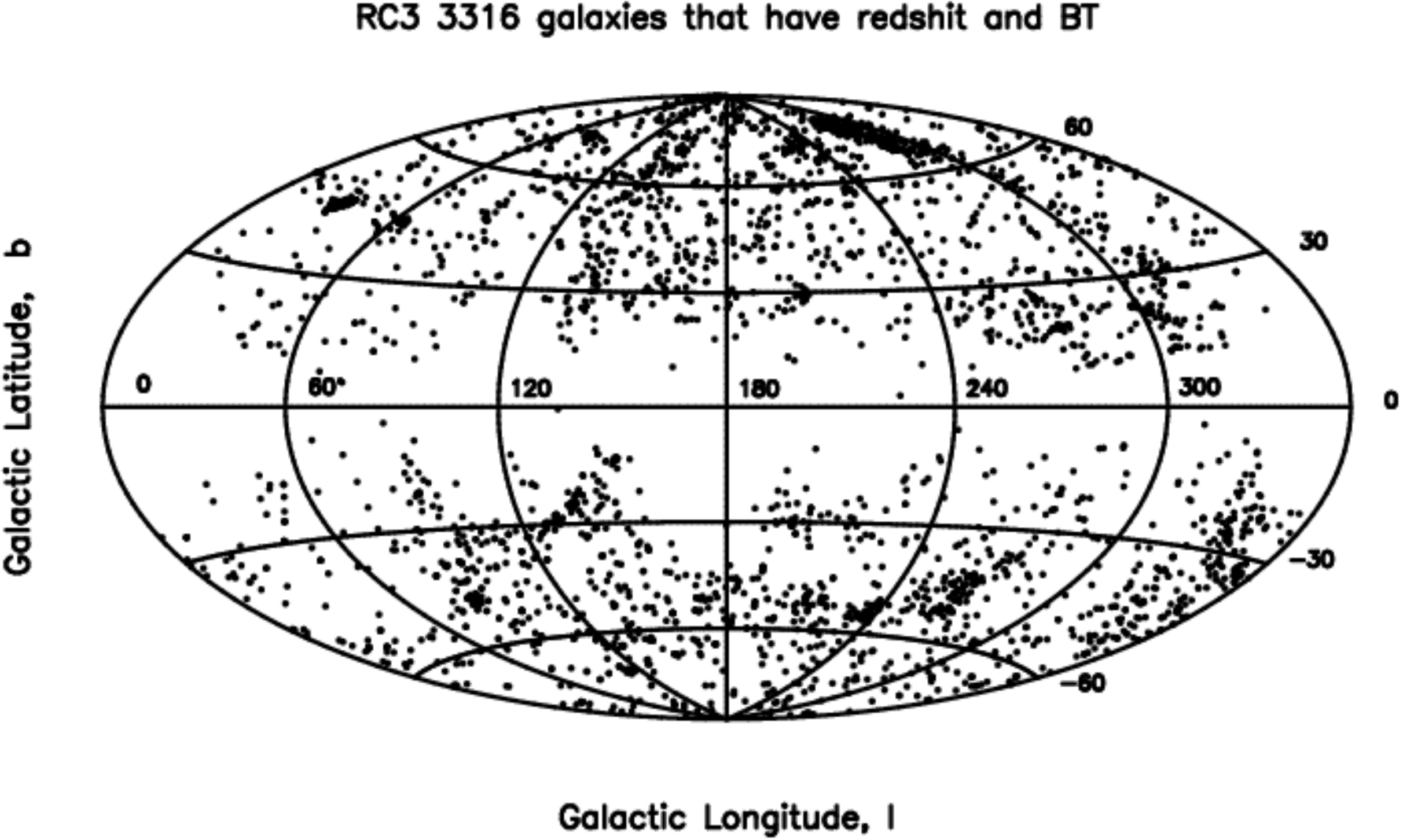}
\end {center}
\caption{
  Hammer-Aitoff  projection in galactic coordinates 
  of 3316  galaxies  
  in the  RC3  which have BT and redshift.
}
          \label{rc3_all}%
    \end{figure*}

\begin{figure*}
\begin{center}
\includegraphics[width=10cm]{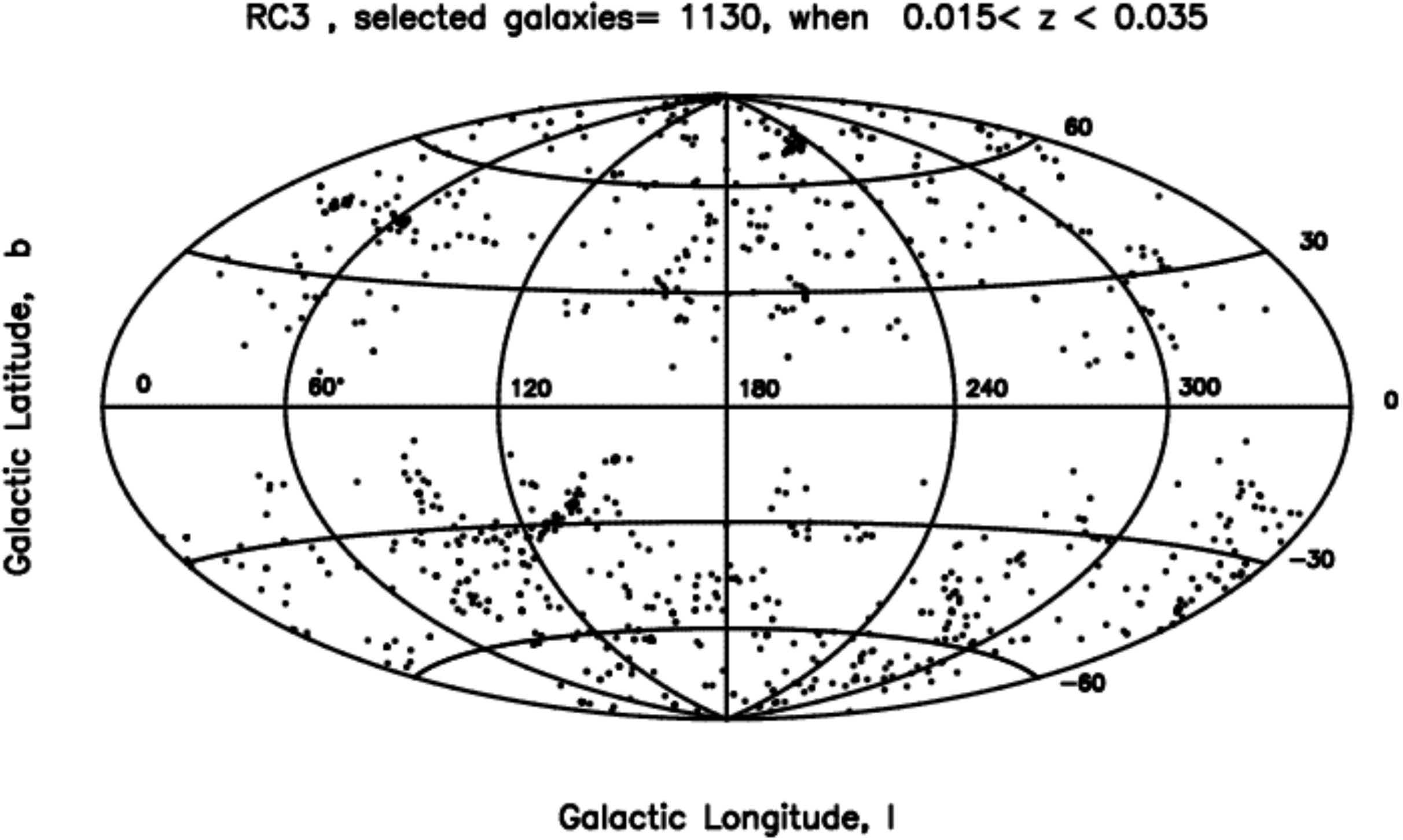}
\end {center}
\caption{
Hammer-Aitoff  projection  
in galactic coordinates 
(observational counterpart 
of  $V_s(2,3)$ ) 
  of 1130  galaxies  
 in the RC3  which have BT and  $0.015 < z < 0.035$.
}
          \label{rc3_z}%
    \end{figure*}

\begin{figure*}
\begin{center}
\includegraphics[width=10cm]{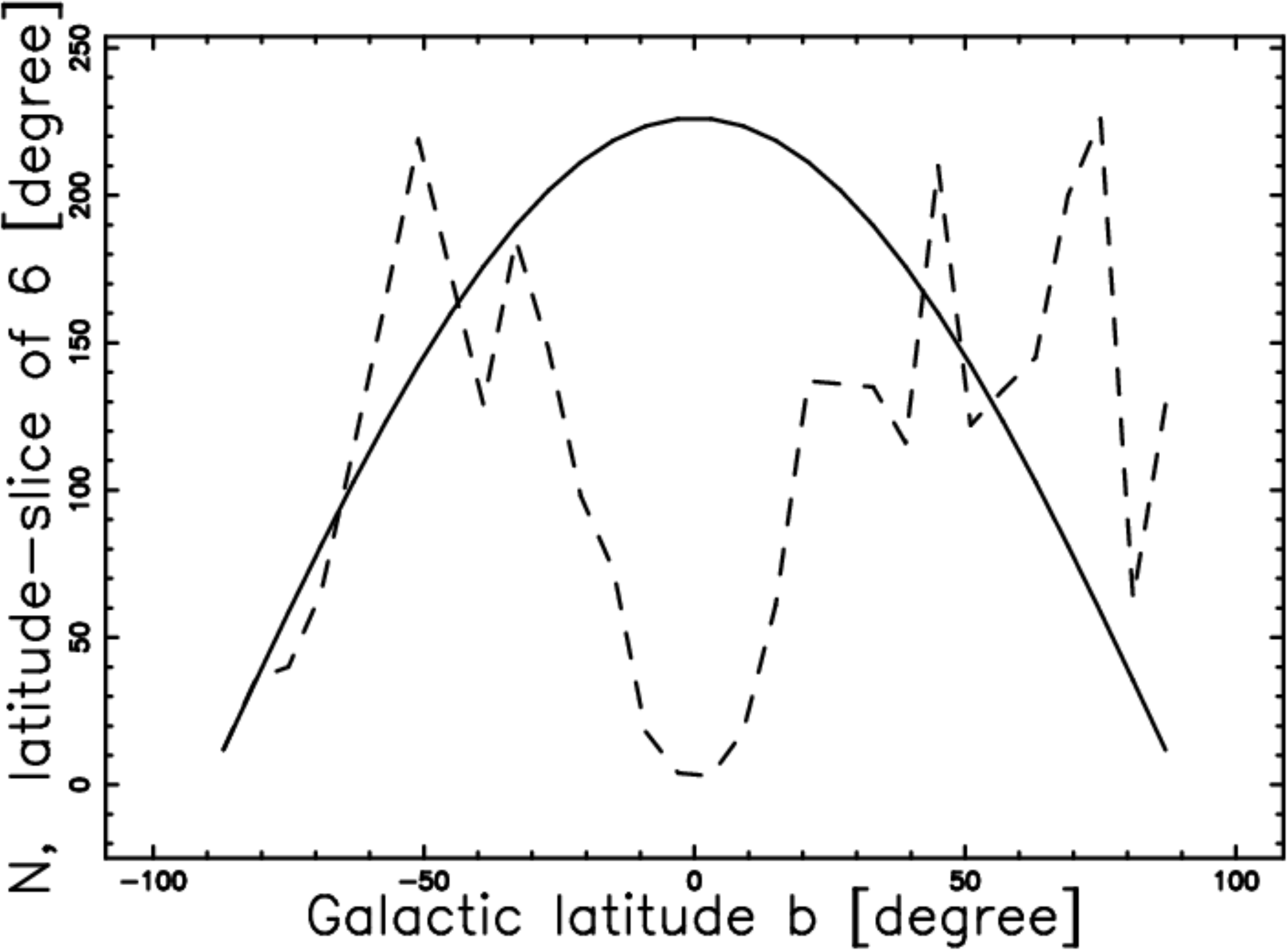}
\end {center}
\caption{
The galaxies  
in the  RC3  which have BT and redshift
are organized in  frequencies versus  galactic latitude  $b$
(dashed line).
The theoretical fit  represents  
$N_i$ ( full line).
}
          \label{teta}%
    \end{figure*}

This catalog    attempts to be reasonably complete for galaxies
having apparent diameters larger than 1 arcmin at the D25
isophotal level and total B-band magnitudes BT,
brighter than about 15.5, 
with a redshift not in excess of 15000 km/s.
All the galaxies in  the
RC3 catalog  which have redshift and BT  are reported
in Figure~\ref{rc3_all}.
Figure~\ref{rc3_z}  reports the RC3 galaxies
in a given window in $z$.
We now test the concept  of an   isotropic universe.
This can be by  done by  plotting  the  number of galaxies  
comprised in
a slice of $360^{\circ}$ in galactic longitude  versus  
a variable number $\Delta b$ in galactic latitude, 
for example $6^{\circ}$.
The number of galaxies in the  RC3 versus galactic latitude is 
plotted  in Figure~\ref{teta}.

The solid angle $d\Omega$ in spherical  coordinates 
(r,$\theta , \phi $) is
\begin{equation}
d \Omega = \sin (\theta) d\theta d \phi \quad .
\end {equation}
In a slice of     $360^{\circ} \times \Delta b $  
the  amount of solid angle,   $\Delta \Omega$,   is 
\begin{equation}
\Delta \Omega
 =  2\pi \bigl (
 (\cos (90^{\circ}) - \cos ( b + \Delta b  ) )   
-(\cos (90^{\circ}) - \cos ( b   ) )
         \bigr )   
~~\mathrm {steradians} 
\label{solid}
\quad .
\end {equation}
The  approximate number  of galaxies  in each  slice 
can be found through the following  approximation.
Firstly, we find the largest value  of the  frequencies 
of galaxies, $F_i$,
versus  $b$, $max(F_i)$
 where the index $i$ denotes  a class in
latitude.
We therefore find  the largest value  of  $\Delta \Omega_i$,
$max(\Delta \Omega_i)$.
The introduction of  the multiplicative  factor $M$
\begin{equation}
M  =  \frac {max(F_i)}{ max(\Delta \Omega_i)}
\quad  ,
\end{equation}
obtains  the following theoretical evaluation of the 
number of galaxies $N_i$ as a  function of the latitude,
\begin{equation}
N_i =  M \times \Delta \Omega_i
\quad .
\label{nteorico}
\end{equation}
This number, $N_i$,  as a function of $b$ is plotted 
in Figure~\ref{teta}.

\begin{figure*}
\begin{center}
\includegraphics[width=10cm]{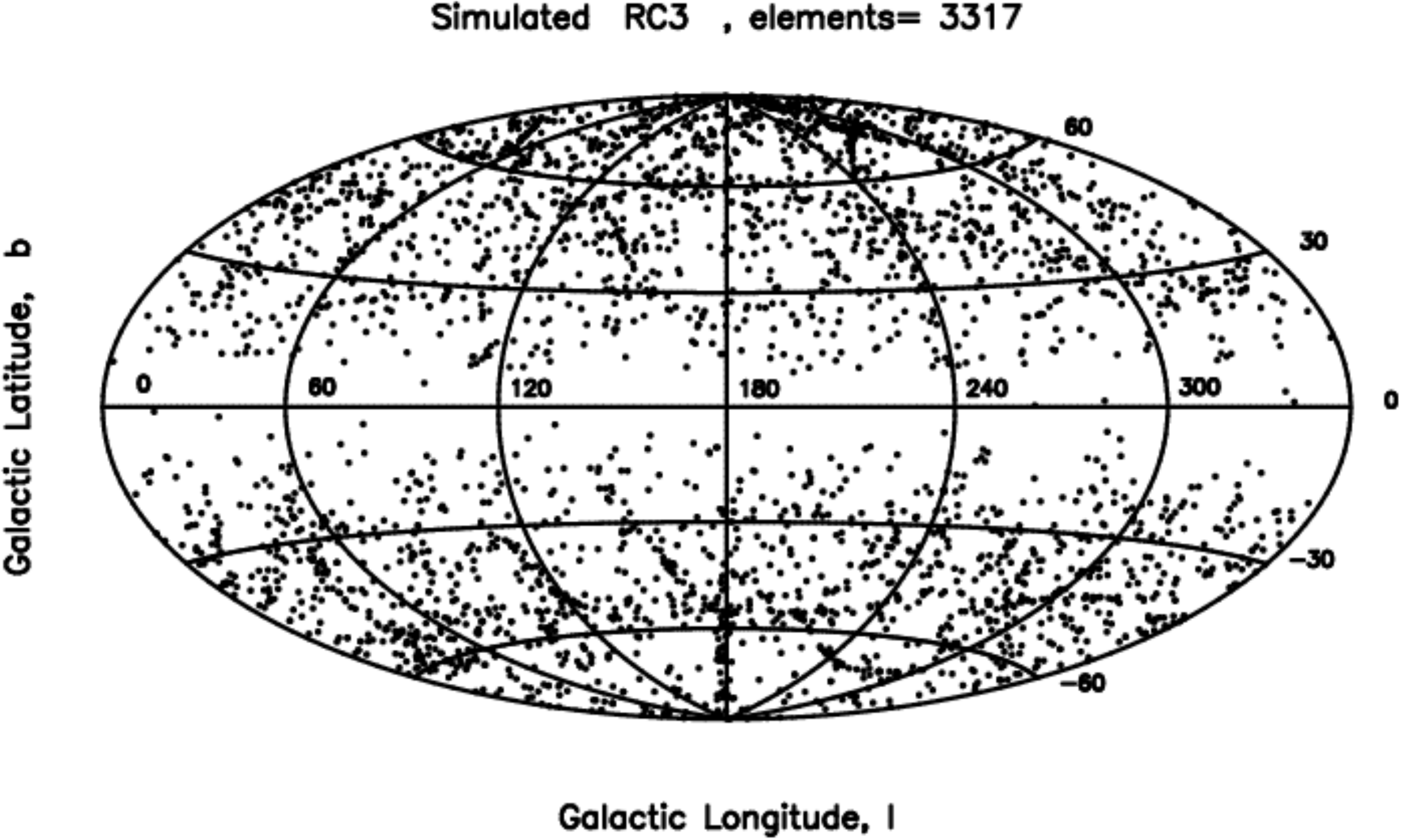}
\end {center}
\caption{Hammer-Aitoff  projection
  of 3317   pixels  belonging to a
  face of an irregular  Voronoi Polyhedron.
  The  {\it Zone of Avoidance}  at the galactic plane  
  follows Figure~\ref{rc3_all}.
  This plot simulates the RC3 galaxies 
  which have BT and redshift.
}
          \label{mix_rc3}%
    \end{figure*}

\begin{figure*}
\begin{center}
\includegraphics[width=10cm]{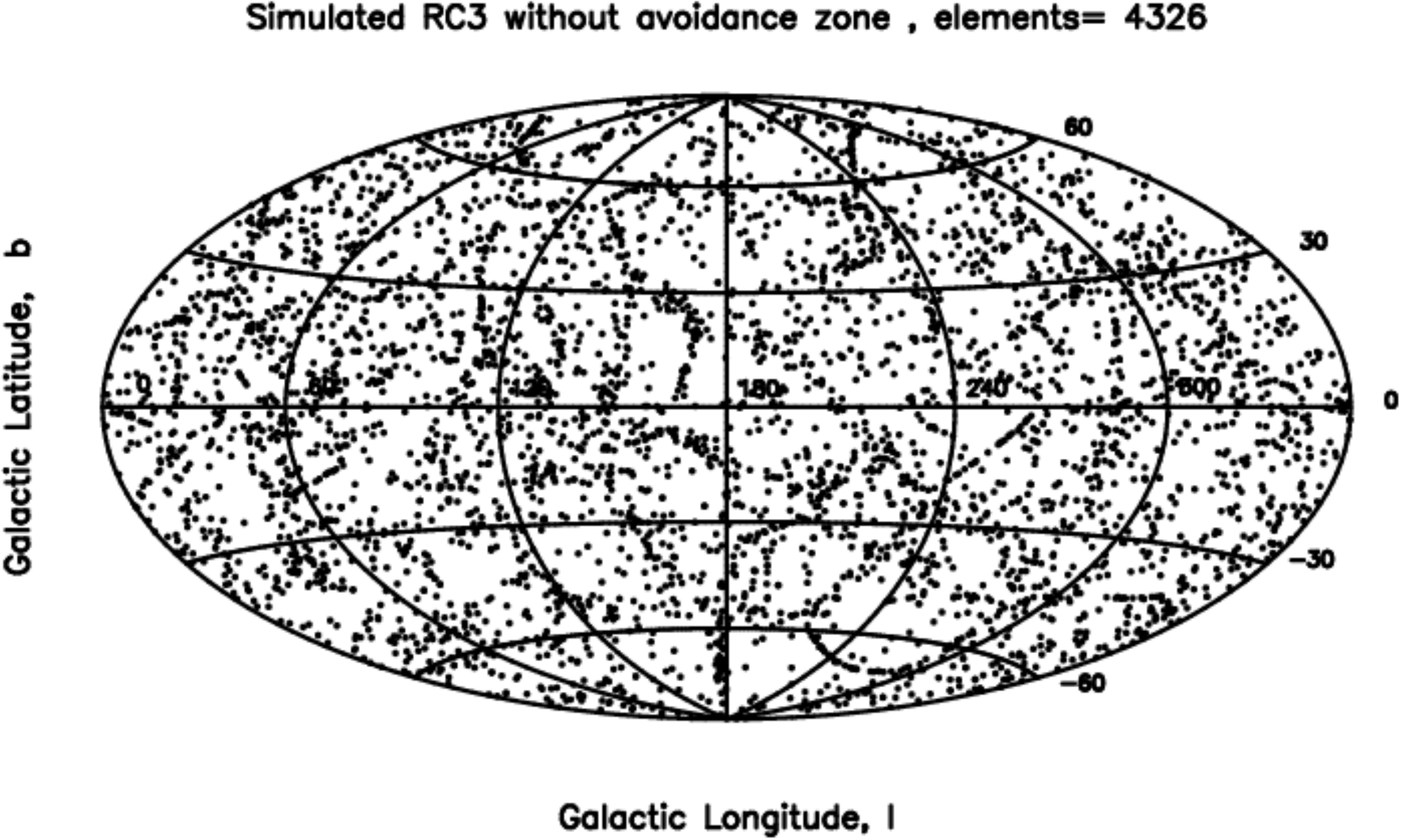}
\end {center}
\caption{Hammer-Aitoff  projection
  of  4326  pixels  belonging to a
  face of an irregular  Voronoi Polyhedron.
  This plot simulates the RC3 galaxies 
  which have BT and redshift but 
  the  {\it Zone of Avoidance}  at the galactic plane  
  is absent.
}
          \label{noavoid_rc3}%
    \end{figure*}

The simulation of this overall sky survey  can be done
in the following way:
\begin {itemize}
\item The pixels  belonging to the faces of  irregular
      polyhedron are  selected according 
      to the distribution in
      $z$ of the galaxies in the RC3 catalog  
      which have redshift and BT.
\item A second operation selects  the  pixels   according 
      to the distribution in latitude in the RC3 catalog,
      see  Figure~\ref{mix_rc3}.
\item In order to simulate  a theoretical  distribution of objects
      which represent the RC3 catalog without the {\it Zone of Avoidance} 
      we made a  series of $6^{\circ}$ slices in latitude 
      in the RC3 catalog, selecting   $N_i$  pixels 
      in  each slice, see  Figure~\ref{noavoid_rc3}.
      In order to ensure that  the range in 
       $z$ is correctly
      described  Table~\ref{rc3catalog} reports 
      $z_{min}$, $z_{pos-max} $, $z_{ave} $ and 
      $z_{max}$   which  represent 
      the minimum $z$, the  position in $z$ of the maximum
      in the number of galaxies,
      and  
      the maximum $z$  in the RC3 catalog or  the simulated sample.
\end {itemize}
\begin{table}
\caption { Real and  Simulated data without the {\it Zone of Avoidance} 
   in the RC3 catalog.   }
 \label{rc3catalog}
 \[
 \begin{array}{ccc}
 \hline
 \hline
 \noalign{\smallskip}
~~   &   RC3   & simulation~ without~ the~ {\it Zone~ of~ Avoidance}        \\
 \noalign{\smallskip}
 \hline
 \noalign{\smallskip}
elements    &  3316                 &  4326                    \\
z_{min}     &  5.7  \times 10^{-7}  & 8.9  \times 10^{-3}      \\
z_{pos-max} &  5.6  \times 10^{-3}  & 8.9  \times 10^{-2}     \\
z_{ave}     &  1.52 \times 10^{-2}  & 7.96  \times 10^{-2}     \\
z_{max}     &  9.4  \times 10^{-2}  & 0.14                    \\
\noalign{\smallskip}
\noalign{\smallskip}
 \hline
 \hline
 \end{array}
 \]
 \end {table}

\subsection{The Eridanus Supervoid}

\label{eridanus}

A void can be defined as the 
empty space between filaments 
in a slice  and 
the typical  diameter has a range of   $[11-50]~Mpc/h$,
see 
\citet{Einasto1994} 
and 
\citet{Einasto1995}.
The probability,   for example,  of  having  a volume
3 times bigger than the average is $3.2~10^{-3}$ 
for PDF~(\ref{rumeni}) when $d=3$ and
$2.1~10^{-3}$ 
for PDF~(\ref{kiangd}) when $d=2.75$.  
Other values of the normalized volume
are reported in Figure~\ref{confronto}.
Particularly  large voids are called super-voids and
have a range of  $[110-163]~Mpc/h$.

Special attention should be paid to  the Eridanus super-void
of 300~$Mpc $  in diameter.
This super-void was  detected by the 
Wilkinson Microwave Anisotropy Probe (WMAP), see 
\citet{Vielva2004}   
\citet{Cruz2005} 
and was  named {\it Cold Spot}.
The  WMAP measures the temperature fluctuations
of the cosmic microwave background (CMB).
Later on the radiostronomers  confirmed the largest   void
due to the  fact that the density of radio sources 
at 1.4 GHz  is anomalously  low
in the direction of the {\it Cold Spot}, see 
\citet{Rudnick2007} 
and 
\citet{McEwen2008}.
The standard statistics of the 
Voronoi  normalized volume distribution in 3D covers the range
$[0.1-10]$.
In the case of a   Eridanus supervoid the normalized volume 
is $\approx~\frac{300}{27}=1.37~10^3$ and the connected
probability of  having  such a supervoid is $1.47~10^{-18}$ 
when the Ferenc \& Neda  function
with $d=3$, formula ~(\ref{rumeni}),
is used and $\approx 0 $ when the Kiang function
with $d=2.75$, formula (\ref{kiangd}) is
used. 
Due to this low probability of  having  such a 
large normalized volume  we 
 mapped a possible
spatial distribution  of the  SDSS-FIRST 
(the Faint Images of the Radio Sky at Twenty cm survey)
sources
with complex radio morphology 
 from
the theoretical distribution of galaxies belonging to the  RC3.
The fraction of galaxies belonging 
to the 2dFGRS   detected as  SDSS-FIRST sources 
with complex radio morphology 
is less than $10\%$  according to  Section 3.8 
in \citet{Ivezic2002}.
We therefore introduced a  probability, $p_{rs}$,
that a galaxy is a radio source.
The  number of  SDSS-FIRST sources $N_{rs}$  in the RC3 which 
are  SDSS-FIRST sources 
with complex radio morphology is 
\begin{equation}
N_{rs} = p_{rs} * N_g
\quad ,
\end{equation}
where $N_g$ is the number of galaxies 
in the theoretical RC3.

\begin{figure*}
\begin{center}
\includegraphics[width=10cm]{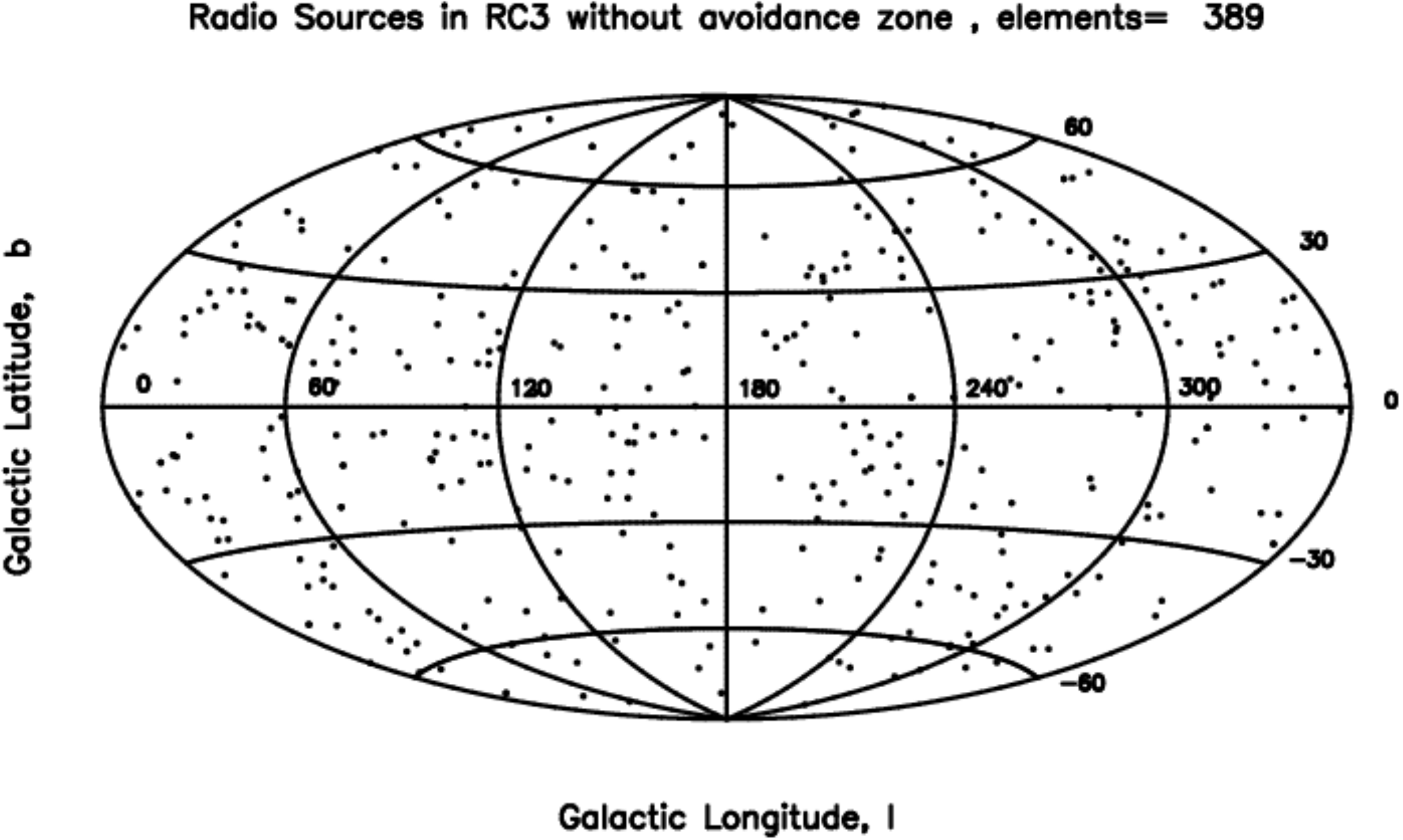}
\end {center}
\caption{
  Hammer-Aitoff  projection
  of the  
  SDSS-FIRST sources
  with complex radio morphology 
  belonging to  the
  RC3, $p_{rs}=0.09$.
  Other parameters as in Figure~\ref{noavoid_rc3}.
}
          \label{rc3_radio}%
    \end{figure*}
From a visual inspection of Figure~\ref{noavoid_rc3} 
and  Figure~\ref{rc3_radio} it is possible
to conclude that the voids increase in size when 
 radiogalaxies which  are a subset of the galaxies
are considered.

\section{The  correlation function for galaxies}
\label{sec_corr}
Galaxies have the tendency to be grouped in clusters 
and a typical measure is the computation of the
two-points  correlation
function for galaxies, see \citet{Peebles1993,Peebles1980}.
The correlation function can be computed in
two ways: a local analysis  in the range
$[0-16] Mpc/h$ and an extended analysis in the range 
$[0-200] Mpc/h$ .

\subsection { The local analysis}

A first  way  to describe the degree  of clustering 
of galaxies is  the two point  correlation  function
$\xi _{GG} (r)$, usually presented in the form
\begin{equation}
 \xi _{GG} = ({r \over r_G})^{-\gamma_{GG }}  
\quad ,
\end{equation}
where  $\gamma_{GG }$=1.8  and  $r_G = 5.77h^{-1} Mpc$   
(the correlation length) when the range 
$0.1 h^{-1} Mpc < r < 16 h^{-1} Mpc$ is considered,
see 
\citet{Zehavi_2004} 
where 118149 galaxies were
analyzed.

In order to compute the correlation function,  
two volumes were compared: one containing 
the little cubes belonging to a face, the other containing 
a random distribution of points. 
From an analysis of the distances of pairs,
the minimum and maximum   were computed
and  $ n_{DD}(r)$   was obtained, 
where $n_{DD}(r)$ is the number of pairs 
of galaxies with separation
within the interval $[r-dr/2, r+dr/2]$.
A similar procedure was applied to the random elements
in the same volume with the same number of elements
and  $n_{RR}(r)$ is the number of
pairs of the Poissonian Process.
According to formula~(16.4.6) in~\citet{coles} 
the correlation function is:  
\begin{equation}
 \xi _{GG} (r) = \frac { n_{DD}(r) } {n_{RR}(r)} -1 \quad . 
\end {equation}
To check whether  $\xi_{GG}$  obeys  a power law or
not we used a simple linear regression test with the formula:
\begin{equation}
 Log \,  \xi _{GG} = a + b\; Log \; r    \quad ,
\end {equation}
which allows us  to compute  $r_G = 10 ^{-a/b}$ and  
$\gamma_{GG }$=-b.

\begin{figure*}
\begin{center}
\includegraphics[width=10cm]{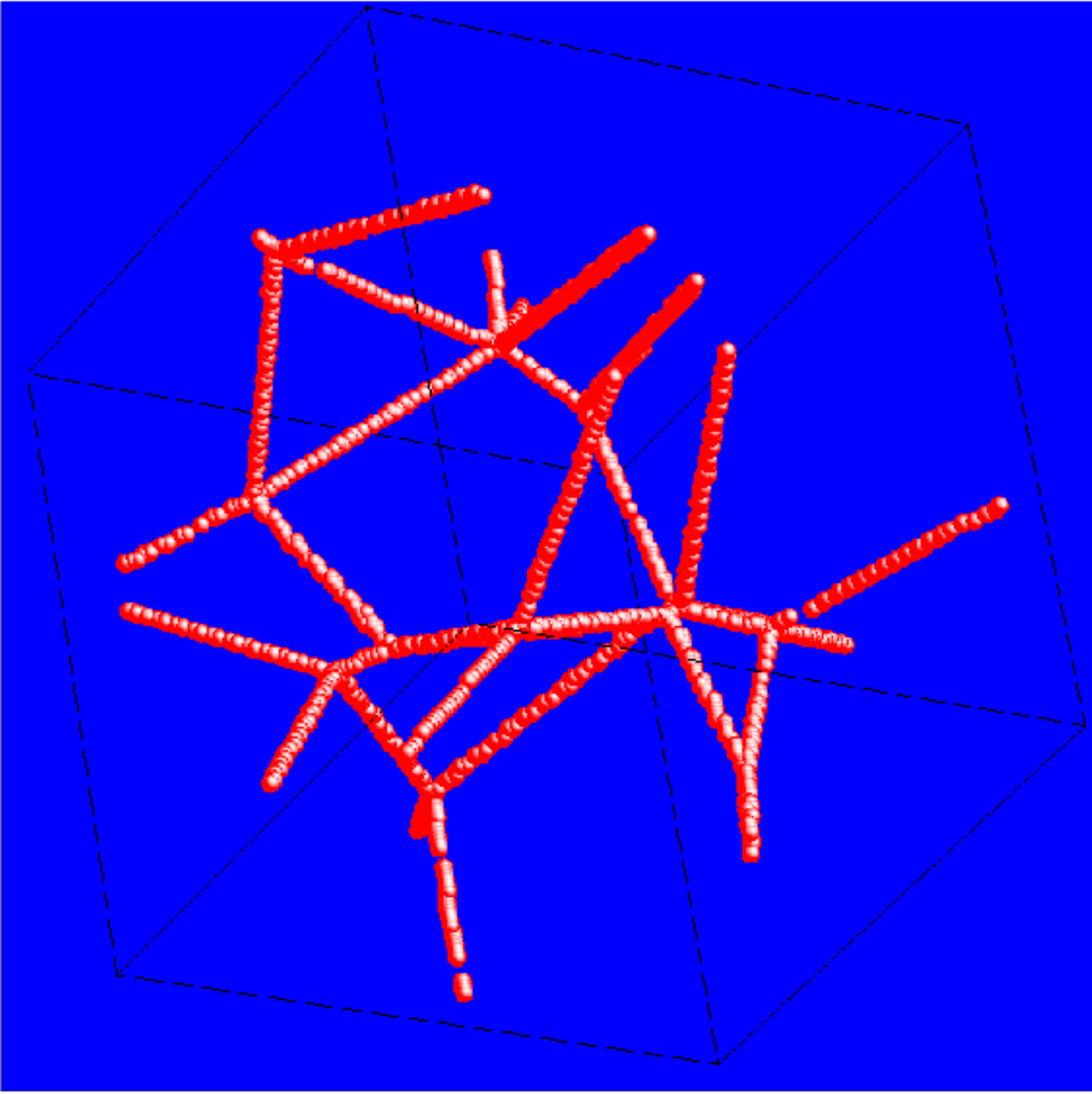}
\end {center}
\caption{
3D visualization of the 
edges of the Poissonian Voronoi--diagram.
The  parameters
are      $ pixels$= 60, $ N_s   $   = 12, 
         $ side  $   = 96.24 $Mpc$, 
         $h=0.623$  and    $ amplify$= 1.2.}
          \label{spigoli3d_sb}%
    \end{figure*}

\begin{figure*}
\begin{center}
\includegraphics[width=10cm]{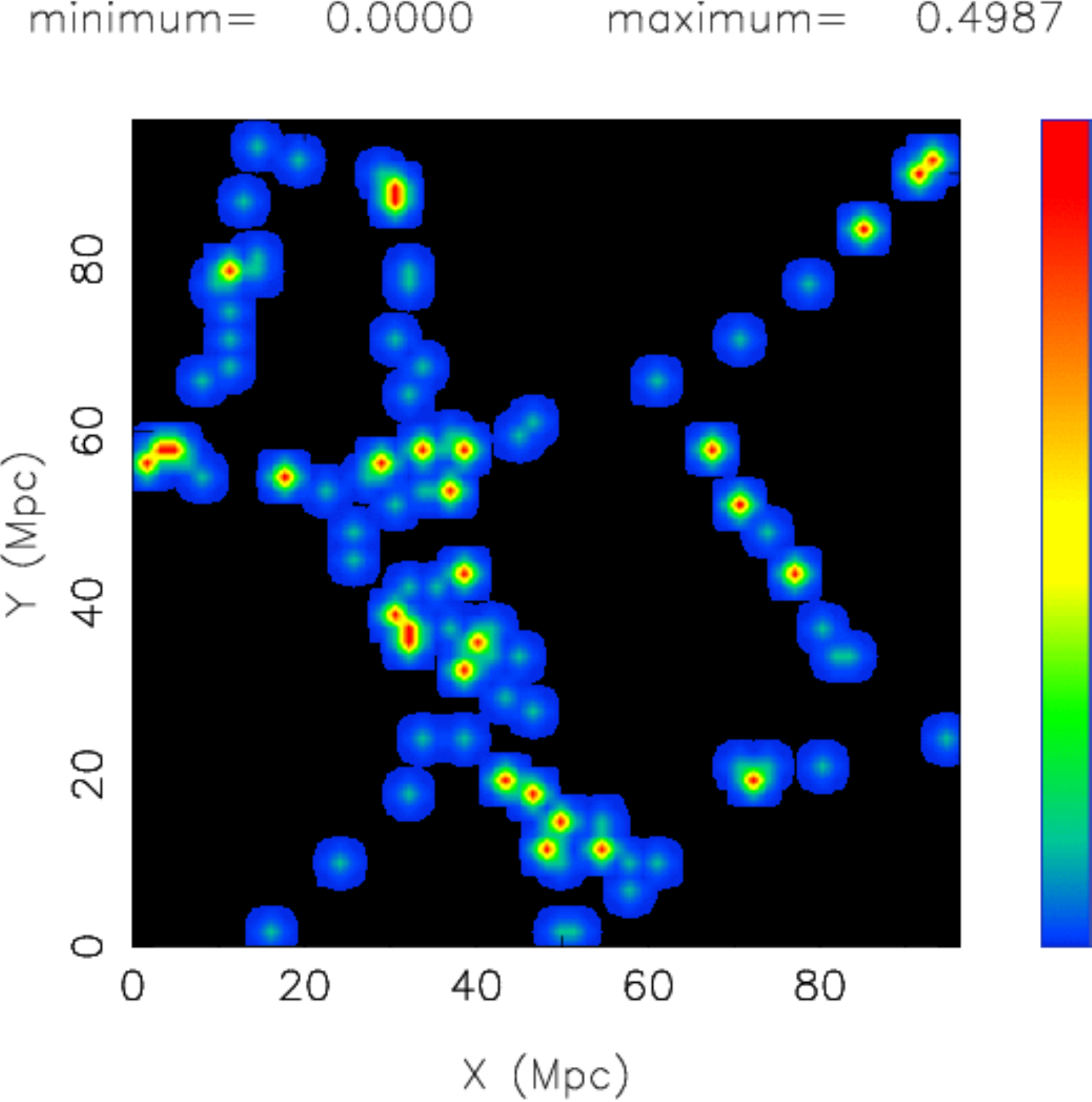}
\end {center}
\caption{
Cut  in the middle of the 3D grid $P(i,j,k)$
which represents a theoretical 2D map 
of the probability of having  
a galaxy. 
The  Voronoi parameters are the same as in  
Figure~\ref{spigoli3d_sb} and $\sigma=0.8 Mpc$.
The X and Y units are in Mpc.
        }
    \label{probability2d}
    \end{figure*}

We now outline the method that allow us 
to compute the correlation
function  using the concept of  thick faces, 
see \citet{zaninetti95}.
A practical implementation is to consider a decreasing probability
of  having  a galaxy  in the direction  perpendicular to the face.
As an example we assume  a  probability, 
$p(x)$, of  having  a galaxy 
outside the face distributed as a Normal (Gaussian) 
distribution
\begin{equation}
p(x) = 
\frac {1} {\sigma (2 \pi)^{1/2}}  \exp {- {\frac {x^2}{2\sigma^2}}} 
\quad  ,
\label{gaussian}
\end{equation}
where $x$ is the distance in $Mpc$ from the face and $\sigma$ 
the standard deviation in $Mpc$.
Once the complex 3D behavior of the faces of the Voronoi
Polyhedron is set up  we can memorize 
such a probability on a 3D grid $P(i,j,k)$ 
which  can be found in the following way  
\begin{itemize}
\item In each lattice point $(i,j,k)$ we search for  
      the nearest  element
      belonging to a Voronoi face. The probability of having  
      a galaxy
      is therefore computed according to formula~(\ref{gaussian}).
\item    A number of galaxies, $N_G=n_* \times side^3$
         is then inserted in the box; 
         here $n_*$ represents the   density of  galaxies 
\end{itemize}
Figure~\ref{spigoli3d_sb} visualizes  
 the edges belonging  to the Voronoi 
diagrams and Figure~\ref{probability2d} represents 
a cut in the middle of the probability, $P(i,j,k)$, 
of having  a galaxy to a given 
distance from a face.

\begin{figure*}
\begin{center}
\includegraphics[width=10cm]{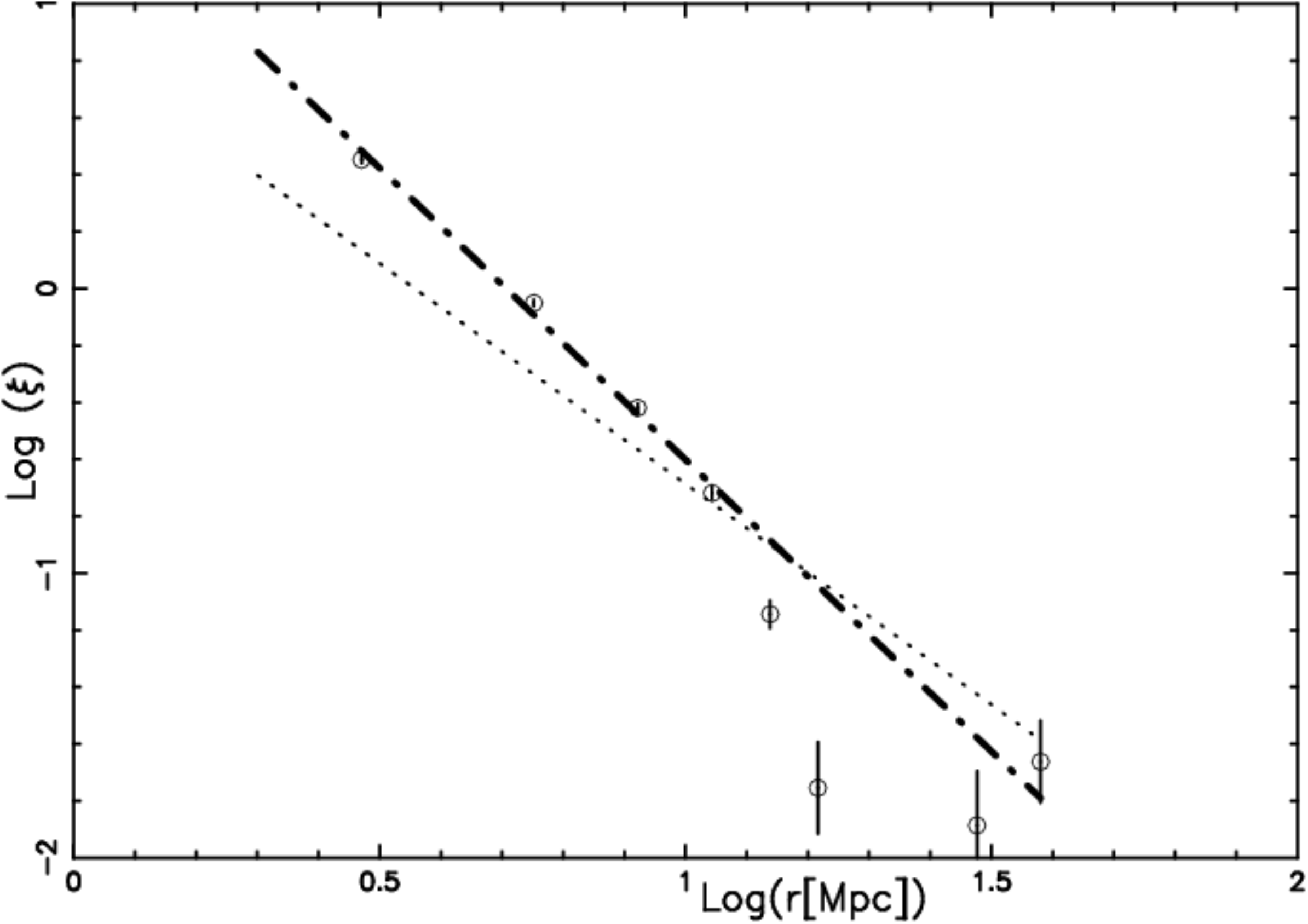}
\end {center}
\caption 
{
The logarithm  of the correlation function is  visualized 
through   points with their uncertainty (vertical bar),
the asymptotic behavior  of the correlation 
function  $\xi_{GG}$  is reported as  dash-dot-dash line;
in our simulation $\gamma_{GG }$=2.04  and  $r_G$ = 5.08 Mpc.
The standard value 
of the correlation function 
is reported  as a  dotted line; 
from the point of view of the observations 
in average $\gamma_{GG }$=1.8  and  $r_G$ = 5 Mpc.
Parameters  of the simulation  as in Figure~\ref{spigoli3d_sb}.
 }
          \label{correlation}%
    \end{figure*}

A typical result of the simulation is reported  in 
Figure~\ref{correlation}  where the center of the 
smaller box in which  the correlation function
is computed   is the  point belonging 
to a face nearest to 
the center of the big box.

From an  analysis of Figure~\ref{correlation}
we can deduce that the correlation 
function $\xi_{GG}$ of the simulation has a behavior  
similar to the standard one.
Perhaps  the value  $r_G$ is a simple measure 
of the face's thickness, $\Delta R_F$.
From this point of view on adopting  a standard value  of the
expanding shell thickness, $\Delta R$ = $\frac{R}{12}$ 
and assuming that the thickness of the shell is made 
by the superposition of two expanding shells
the following is obtained
\begin{equation}
\Delta R_F  \approx  \frac{R}{6} 
\approx \frac{\overline{D^{obs}}}{h\,12} = 3.62~Mpc
\quad ,
\end{equation}
where $h=0.623$ has  been used.
 The 
correlation dimension $D_2$, see~\citet{Jones2005}, 
is connected with the exponent $ \gamma$  
through the relation:
\begin{equation}
 D_2  = 3 - \gamma      
\quad .
\end{equation}
Here there is the case 
in which the  mass M(r) increases
as   $r^{1.2}$,  in the middle  of a one dimensional
structure ( $M(r) \propto r$) and a two dimensional
sheet ( $M(r) \propto r^2$), see \citet{coles}.
In this paragraph the dependence of the correlation
function  on  the magnitude is   not  considered.

\subsection { The extended analysis}

\begin{figure*}
\begin{center}
\includegraphics[width=10cm]{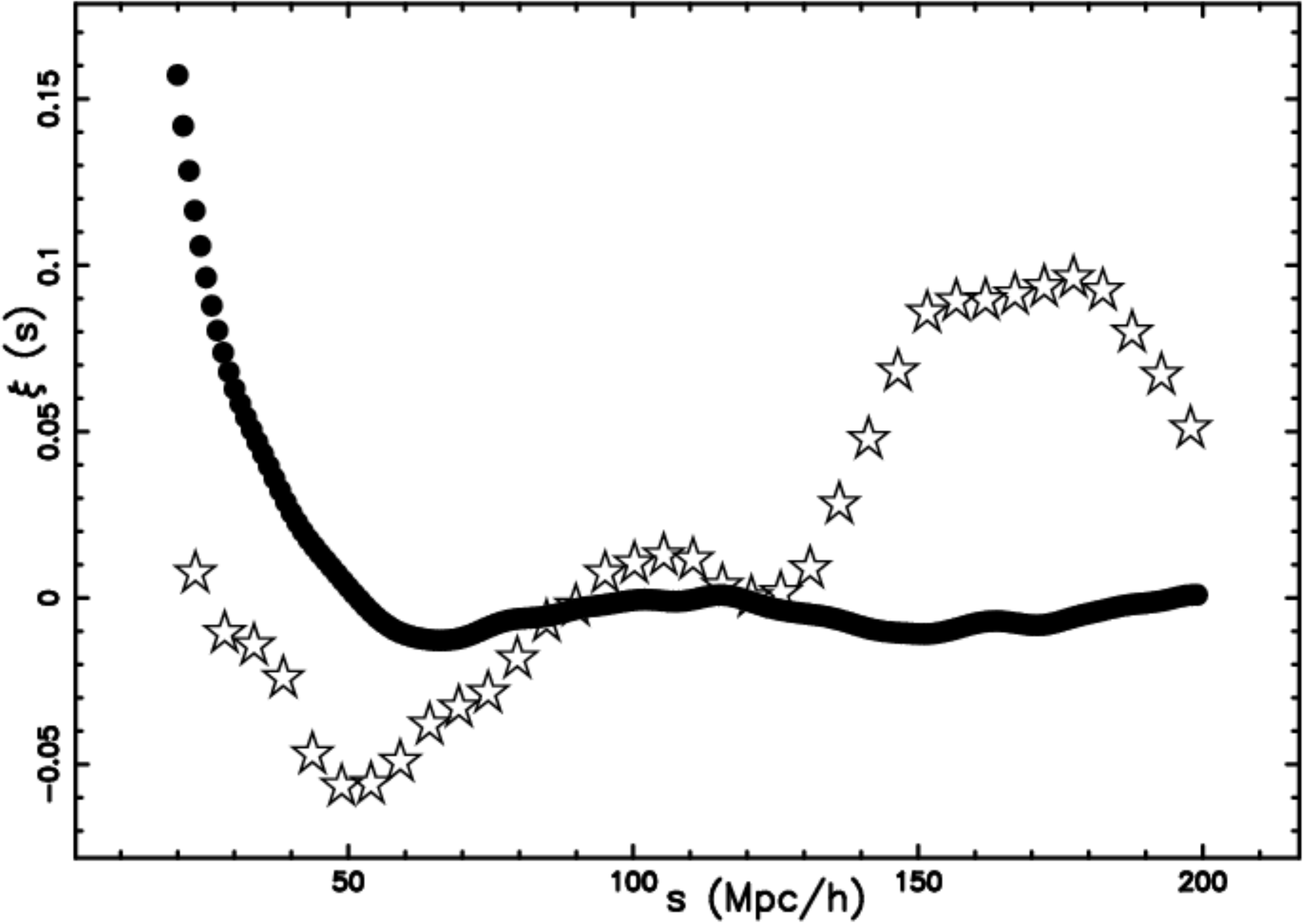}
\end {center}
\caption{
Redshift-space correlation function 
for  the 2dFGRS sample limited 
at $z=0.12$ as given by  our code          ( empty stars )
 and the results  of \citet{Martinez2009}  ( full points )
for 2dFVL.  
The covered range is $[40-200] Mpc/h$ .
        }
    \label{2df_our_martinez}
    \end{figure*}

A second definition of the correlation function 
takes account of the Landy-Szalay border correction,
see \citet{Szalay1993},
\begin{equation}
 \xi _{LS} (s) = 1
+ \frac { n_{DD}(s) } {n_{RR}(s)} 
- 2  \frac { n_{DR}(s) } {n_{RR}(s)} 
 \quad . 
\end {equation}
where $n_{DD}(s)$,  
$n_{DD}(s)$  and  $n_{DR}(s)$
are the number of  galaxy-galaxy ,random-random 
and galaxy-random pairs having distance $s$,
see \citet{Martinez2009}.
A random catalog of galaxies in polar coordinates
can built by generating 
a first random number $\propto~z^2$ in the z-space
and a second random angle in the interval $\bigl [0,75 \bigr ]$. 
A test of our code for the correlation 
function versus  a more sophisticated code 
is reported in Figure~\ref{2df_our_martinez}
for the 2dFVL volume limited (VL) sample,
where the data available at the   
Web site http://www.uv.es/martinez/ 
have been processed.

\begin{figure*}
\begin{center}
\includegraphics[width=10cm]{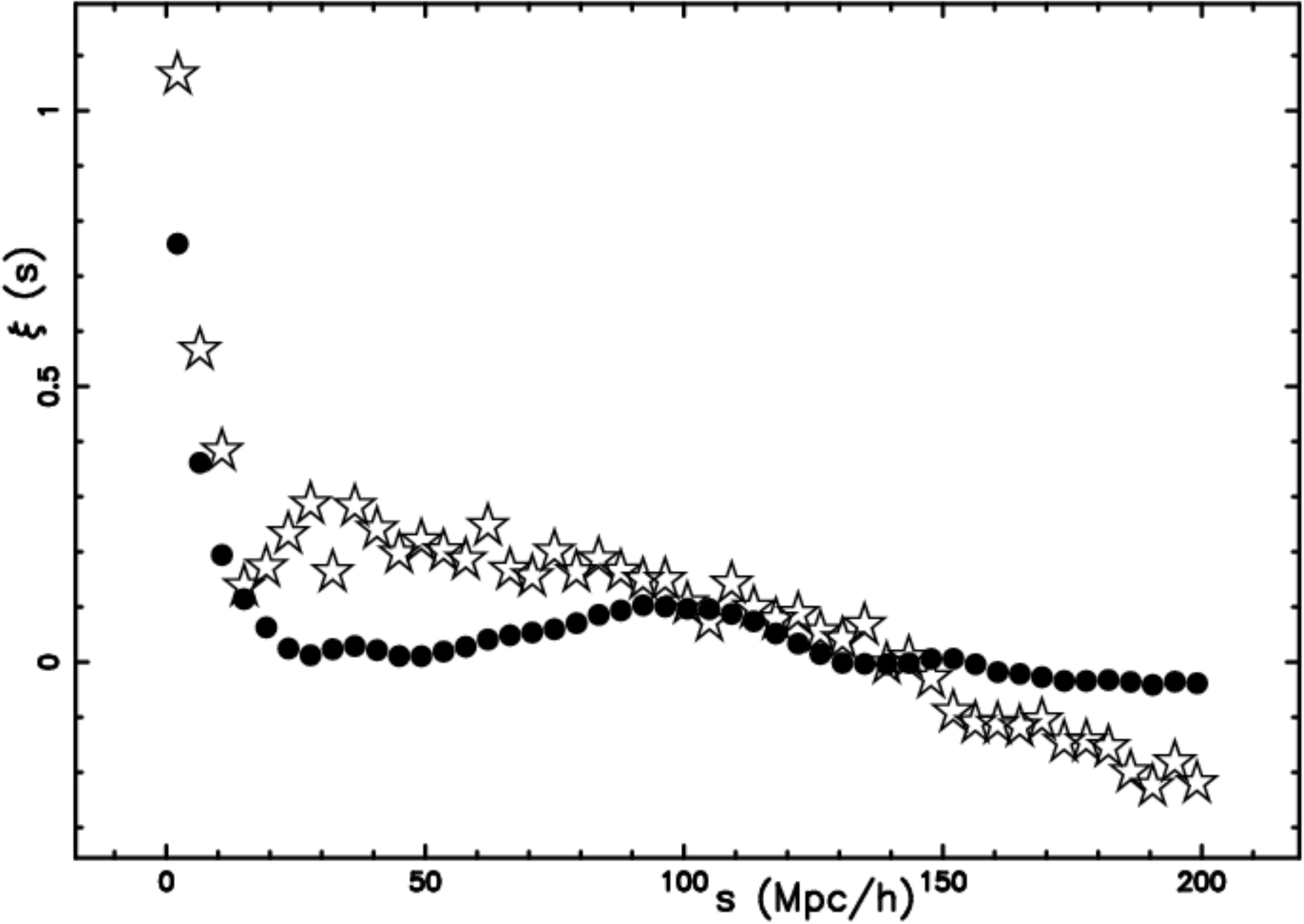}
\end {center}
\caption{
Redshift-space correlation function 
for  the 2dfGRS sample ( empty stars )
 and the Voronoi sample( full points ).  
The covered range is $[40-200] Mpc/h$.
        }
    \label{correlation_due}
    \end{figure*}

The pair correlation function for the vertexes of the
Poissonian Voronoi Polyhedron
 presents a typical damped oscillation,
see Figure 5.4.11 in \citet{okabe}, Figure 2 in 
\citet{Martinez2009}
and Figure 3 in \citet{Heinrich2008}.
Here conversely : (a) we first consider a set of objects belonging
to the faces of the irregular Polyhedron
; (b) we extract from the previous set a subset which follows 
the photometric law and then  
we compute the pair correlation function. 
The difference between our model and the model in 
\citet{Martinez2009}  
for 2dFVL
can be due to the luminosity color segregation 
presents in 2dFVL but not in our Voronoi type model. 
A typical  result is  reported in Figure~\ref{correlation_due}
where it is possible to find the correlation function
of 2dfGRS   with astronomical 
data as reported in Figure~\ref{2df_all}
as well  as the correlation function of the Voronoi 
network with simulated data as reported
in Figure \ref{voro_2df_cones}.

A careful analysis of Figure~\ref{correlation_due}
allows us to conclude that the behavior of the correlation
function is similar for the astronomical data as well
as the simulated Voronoi-data.
The oscillations 
after 100 $Mpc$ are classified as 
acoustic,  \citet{Eisenstein2005}.

\section{Summary}

{\bf Photometric maximum}

The  observed number of galaxies in a given  solid angle  
with a chosen   flux/magnitude  
versus the redshift presents 
a maximum  that is a function of the flux/magnitude.
From  a theoretical  point of view,   
the  photometric  properties of the galaxies 
depend on  the chosen law  for the luminosity function.
The  luminosity function  here adopted (the 
Schechter function)  
predicts a maximum 
in the theoretical number of galaxies  as a  
function of the redshift
once the apparent flux/magnitude is fixed.

The theoretical  fit  representing  the number 
of galaxies as a function 
of the redshift can be compared with the 
real number of galaxies
of the 2dFGRS which  is theory-independent.
The superposition of theoretical  and observed fit  
is satisfactory   and the $\chi^2$ has  been computed,
see Figure~\ref{maximum_flux}.
The position of the maximum in the number of galaxies 
for different magnitudes is a function of the redshift 
and in the interval 
$15 < bJmag< 18.5 $
the comparison between observed and theoretical data is
acceptable, see Figure~\ref{zeta_max_flux}.
Particular attention should be paid
to the  Malmquist bias and to equation~(\ref{range}) 
that  regulate  the upper value of the 
redshift that defines the complete sample.

\noindent
{\bf 3D Voronoi  Diagrams}
The intersection between a plane and the 3D Voronoi faces
is  well known as 
$V_p(2,3)$.
The intersection between a slice of a given opening angle,
for example $3^{\circ}$, and the 3D Voronoi faces
is less known  and has  been developed  in 
Section~\ref{faces}.
This intersection can be calibrated 
on the astronomical data  once the number of 
Poissonian 
seeds   is such that the largest observed void 
matches  the largest Voronoi volume.
Here  the largest observed void is 
2700 $Km/sec$ and  in order to simulate, for  example,
the 2dFGRS,   137998  Poissonian seeds  were inserted 
in a volume of
$(131908~Km/sec)^3$.
The intersection between a sphere 
and the 3D Voronoi faces
represents a new way to visualize 
the voids in the distribution of galaxies,
see Section~\ref{faces}.
In this  spherical  cut the intersection 
between a sphere
and the 3D Voronoi faces
is no longer represented 
by  straight lines but by curved lines 
presenting  in some cases a  cusp behavior at the 
intersection, see Figure~\ref{aitof_sphere}.
In line of principle the spatial distribution
of  galaxies at a given redshift  should
follow   such  curved lines.

\noindent
{\bf Statistics of the voids}
The  statistical properties of the voids 
 between galaxies 
can be well described by the volume  distribution 
of the Voronoi Polyhedra.
Here two distributions of probability were carefully compared:
the old Kiang function here parametrized 
as a function of the dimension $d$  , see formula~(\ref{kiang}),
and the new distribution 
of Ferenc ~\&~ Neda~ , see formula~(\ref{rumeni}), 
which  is a function of the selected 
dimension $d$.
The  probability of  having  voids as large
as the Eridanus super-void was  computed,
see Section~\ref{eridanus}.

\noindent
{\bf Simulations of the catalogs of galaxies}
By combining  the photometric dependence 
in the number of galaxies 
as a function of the redshift with the intersection
between a slice and the Voronoi faces, it  is possible to simulate
the astronomical catalogs  
such as the  2dFGRS, see Section~\ref{cat2dFGRS}.
Other catalogs such as  
the RC3 which covers 
all the sky ( except the   Zone of Avoidance ) 
can be simulated through a given number of 
spherical cuts, for example 25,
with progressive increasing redshift.
This  simulation is visible 
in  Figure~\ref{mix_rc3}  in which the 
theoretical influence   of the 
Zone of Avoidance  has been inserted,
and in Figure~\ref{noavoid_rc3} 
in which the theoretical RC3 without the 
Zone of Avoidance has  been modeled.
Figure~\ref{rc3_radio} reports the subset 
of the  galaxies which  are  radiogalaxies.

\noindent
{\bf Correlation function}
The standard behavior of the correlation function  for  
galaxies in the short range 
$[0-10~Mpc/h]$ 
can be simulated  once 12 Poissonian seeds are inserted 
in a  box of volume
$( 96.24~Mpc/h )^3$ .
In this case the model can be refined
by 
 introducing the concept
of galaxies generated in   a thick face 
belonging to the Voronoi  Polyhedron.
The behavior of the correlation function in the large range 
$[40-200~Mpc/h]$  of the Voronoi   simulations 
of the 2dFGRS  presents minimum variations from  
 the processed astronomical data, see 
Figure ~\ref{correlation_due}.
We now extract a question from the conclusions of 
\citet{Martinez2009} ``Third, the minimum in 
the large-distance correlation functions
of some samples demands explanation: is it really the
signature of voids?''
Our answer is ``yes''. The  minimum in the large scale
correlation function is due to the combined effect 
of the large empty space between galaxies ( the voids )
and to  the photometric behavior of  the number of 
galaxies as  a function of the red-shift.


\appendix

\setcounter{equation}{0}
\renewcommand{\theequation}{\thesection.\arabic{equation}}

\section{The maximum likelihood estimator }
\label{appendix_mle}

The parameter $\widehat z_{crit}$ can be derived through the maximum
likelihood estimator (MLE)
and as a consequence $z=z_{pos-max}$  is easily derived.
The likelihood function is defined as the probability
to obtain a  set of observations 
if given  particular set 
of the distribution parameters,$c_i$,
\begin{equation}
L(c) = f (x_1 \ldots x_n  | c_1 \ldots c_n)
\quad .
\end {equation}
If we assume that the $n$ random variables
are independently and
identically distributed,
then we may write the likelihood function as
\begin{equation}
L(c) = f (x_1 |  c_1 \ldots c_p) \ldots
f (x_n| c_1 \ldots c_p) =
\prod_{i=1}^n f (x_i | c_1 \ldots c_p)
\quad .
\end {equation}

The maximum likelihood estimates for the
$c_i$  are
obtained by maximizing
the likelihood function, L(c).
In the same way, 
we may  find it easier to
maximize  $ln  f(x_i)$,
termed the log-likelihood.
So, for a random sample $z_1 \ldots  z_n$
representing the redshift of the galaxies that fall in
a given interval of flux or magnitude
from a joint distribution in {\it z}  and {\it f} 
for galaxies  adopting the Schechter function 
for the luminosity represented by equation~ (\ref{nfunctionz}), 
the
likelihood function  is
given by
\begin{equation}
L(z_{crit})  =
\prod_{i=1}^n (z_i)^{2\alpha +4} \frac {1}{(z_{crit})^{2\alpha}}
\exp {
-\frac{z_i^2}
      {z_{crit}^2}
}  
\quad ,
\end {equation}
where the constant terms are omitted.
Using logarithms, we obtain the log-likelihood
\begin {equation}
\ln L(z_{crit}) =({2\alpha +4})\sum_{i=1}^n  \ln {z_i}
+  n 2 \mid \alpha \mid \ln (z{crit}) 
-\sum_{i=1}^n 
 \frac{z_i^2     }
      {z_{crit}^2}
\quad .
\end {equation}
Taking the first derivative  with respect to 
$z_{crit}$
equal to zero, we get
\begin {eqnarray}
\widehat  z_{crit} = \sqrt
{
\frac 
{
\sum_{i=1}^n z_i^2 
}
{
n \mid \alpha \mid
}
}
\quad .
\end {eqnarray}
According to equation~(\ref{massimoschecher}) 
$\widehat z_{pos-max}$
is  
\begin{equation}
\widehat  z_{pos-max} = \sqrt
{
\frac 
{
\sum_{i=1}^n z_i^2 
}
{
n \mid \alpha \mid  
}
}  
\sqrt {\alpha +2 }
\label{massimoschecherlike}
\end{equation}

When a joint distribution in {\it z}  and {\it f} 
for galaxies  in the presence of  the ${\mathcal M}-L$
relationship  is considered, see equation~(38) in 
 \citet{Zaninetti2008}, the likelihood function  is
\begin{equation}
L(z_{crit})  =
\prod_{i=1}^n (z_i)^{4+2\frac{c-a}{a}  } 
{
(z_{crit})^{-2\frac{c-a}{a}}
}
\exp^
{
-\bigl (
\frac{z_i^2}
       {z_{crit}^2}
\bigr       )
^{1/a} 
}
\quad ,
\end {equation}
where the constant terms are omitted.
Taking the first derivative  with respect to 
$z_{crit}$
of $\ln L(z_{crit})$
equal to zero, we get
for the ${\mathcal M}-L$
relationship
\begin {eqnarray}
\widehat  z_{crit} = 
\bigl (
\frac 
{
\sum_{i=1}^n z_i^{2/a} 
}
{
n \mid c-a  \mid
}
\bigr )  ^{a/2}
\quad .
\end {eqnarray}
According to equation~(\ref{zmassimomia}) 
$\widehat z_{pos-max}$
for the ${\mathcal M}-L$
relationship
is  
\begin{equation}
\widehat  z_{pos-max} = 
\bigl (
\frac 
{
\sum_{i=1}^n z_i^{2/a} 
}
{
n \mid c-a  \mid
}
\bigr )  ^{a/2}
\bigl (
a +c
\bigr )^{a/2}
\label{massimomialike}
\quad . 
\end{equation}

\section* {Acknowledgements}
I thank the
2dF Galaxy Redshift Survey team for the use of 
Figure \ref{2df_cone}, which is
taken from the image gallery on the 2dFGRS website (see
http://www2.aao.gov.au/2dFGRS).


\end{document}